\documentclass[prl,amsmath,amssymb,aps,superscriptaddress]{revtex4-2}
\usepackage{amsfonts}
\usepackage{stmaryrd}
\usepackage{bbm}
\usepackage{mathrsfs}
\usepackage{tipa}
\usepackage{amssymb}
\usepackage{txfonts}
\usepackage{graphicx}
\usepackage{dcolumn}
\usepackage{epstopdf}
\usepackage[colorlinks,linkcolor=blue,urlcolor=blue,citecolor=blue]{hyperref}
\usepackage{multirow}
\usepackage{subfigure}
\usepackage{url}
\usepackage{upgreek}
\usepackage[utf8]{inputenc}
\usepackage[english]{babel}

\usepackage{color}
\usepackage{graphicx}

\newcommand{\Vect}[1]{\mathbf{#1}} 
\newcommand{\ham}{\hat{H}} 
\newcommand{\hfield}{B} 
\newcommand{\mat}[1]{\mathcal{#1}} 
\newcommand{\transpose}[1]{#1^{\text{T}}} 
\newcommand{\deltagamma}{\gamma}
\newcommand{\idmat}{\mathbf{1}} 
\newcommand{\KitaevAxis}{(\text{K})} 
\newcommand{\subrot}{\mathrm{C}} 
\newcommand{\Jeff}{\tilde{J}} %
\newcommand{\CWtemp}{\Theta} 
\newcommand{\nd}{{\vphantom{\dagger}}} 

\begin{document}


\title{ Field-induced spin continuum in twin-free Na$_3$Co$_2$SbO$_6$ revealed by magneto-THz spectroscopy}

\author{Rongsheng Li}
\altaffiliation{These authors contributed equally to this work.}
\author{Liyu Shi}
\altaffiliation{These authors contributed equally to this work.}
\email{shiliyu@pku.edu.cn}
\author{Yuchen Gu}
\author{Yue Chen}
\affiliation{International Center for Quantum Materials, School of Physics, Peking University, Beijing 100871, China}
\author{Xintong Li}
\affiliation{International Center for Quantum Materials, School of Physics, Peking University, Beijing 100871, China}
\affiliation{Beijing National Laboratory for Condensed Matter Physics, Institute of Physics, Chinese Academy of Science, Beijing 100190, China}

\author{Qiong Wu}
\author{Shuxiang Xu}
\affiliation{International Center for Quantum Materials, School of Physics, Peking University, Beijing 100871, China}

\author{Dong Wu}
\affiliation{Beijing Academy of Quantum Information Sciences, Beijing 100193, China}

\author{Yuan Li}
\email{yuan.li@iphy.ac.cn}
\affiliation{International Center for Quantum Materials, School of Physics, Peking University, Beijing 100871, China}
\affiliation{Beijing National Laboratory for Condensed Matter Physics, Institute of Physics, Chinese Academy of Science, Beijing 100190, China}
\affiliation{Collaborative Innovation Center of Quantum Matter, Beijing 100871, China}

\author{Fa Wang}
\email{wangfa@pku.edu.cn}
\affiliation{International Center for Quantum Materials, School of Physics, Peking University, Beijing 100871, China}
\affiliation{Collaborative Innovation Center of Quantum Matter, Beijing 100871, China}

\author{Tao Dong}
\email{taodong@pku.edu.cn}
\affiliation{International Center for Quantum Materials, School of Physics, Peking University, Beijing 100871, China}
\author{Nanlin Wang}
\email{nlwang@pku.edu.cn}
\affiliation{International Center for Quantum Materials, School of Physics, Peking University, Beijing 100871, China}
\affiliation{Beijing Academy of Quantum Information Sciences, Beijing 100193, China}
\affiliation{Collaborative Innovation Center of Quantum Matter, Beijing 100871, China}

\date{\today}

\begin{abstract}
The honeycomb magnet Na$_3$Co$_2$SbO$_6$ recently emerged as a promising candidate for realizing Kitaev quantum spin liquid with relatively low levels of structural disorder. While the precise spin Hamiltonian remains controversial, the potential existence of a quantum spin liquid or other novel quantum magnetic phases continues to stimulate investigation. Here, we study the temperature and magnetic field-dependent spin excitations of Na$_3$Co$_2$SbO$_6$ on a twin-free single crystal using magneto-terahertz (THz) spectroscopy, focusing on magnetic anisotropy and field-induced unusual phases. We observe a low-energy continuum excitation above $T_N$ and a 0.5 THz (2 meV) spin wave excitation in magnetic order under zero field. Upon applying an in-plane magnetic field, the spin waves transform to a magnetic continuum over an intermediate field range, above which the system enters a spin-polarized state. Crucially, the spin excitation spectra reveal striking anisotropy between the $\textbf{a}$ and $\textbf{b}$ crystallographic axes, demanding description by a strongly anisotropic spin model. These findings establish Na$_3$Co$_2$SbO$_6$ as a model system for investigating field-tunable quantum magnetism and potential spin liquid behavior in highly anisotropic systems.

\end{abstract}

\pacs{Valid PACS appear here}
\maketitle



Quantum spin liquids (QSLs) are exotic states of matter characterized by long-range entanglement, topological order, and fractional excitations~\cite{ANDERSON1973153,balents2010spin,knolle2019field,takagi2019concept,freedman2003topological, motome2020hunting}. The exactly solvable Kitaev model on a honeycomb lattice has inspired significant advances to search for QSLs \cite{KITAEV20062,takagi2019concept,banerjee2017neutron, Hermanns2018, Takagi2019, PhysRevLett.119.127204, Banerjee2018, PhysRevLett.112.207203, Winter2017,Hickey2019, Do2017,Wulferding2020, Bruin2022,Czajka2023}. Despite considerable theoretical progress, experimental realization of Kitaev QSLs in materials remains challenging \cite{Wen2019}. Initial efforts focused on 4$d$/5$d$ transition compounds like iridium oxides and $\alpha$-RuCl$_3$, where strong spin-orbit coupling and electronic correlations were predicted to stabilize Kitaev interactions~\cite{witczak2014correlated, PhysRevB.90.041112, PhysRevLett.105.027204}. However, in all these materials, significant Heisenberg and off-diagonal exchange interactions hinder the emergence of Kitaev QSL states, leading to long-range magnetic order at low temperatures~\cite{Takagi2019, PhysRevLett.105.027204, PhysRevB.84.180407, PhysRevB.96.115103, PhysRevB.96.064430, PhysRevB.99.064425, PhysRevResearch.2.033011}. Current research strategies focus on either identifying materials with suppressed non-Kitaev interactions or employing external parameters to suppress the long-range order \cite{Banerjee2018, PhysRevLett.120.077203, PhysRevB.97.245149}.

Following the first strategy, cobalt-based honeycomb magnets with a high-spin $3d^7$ configuration have been proposed as Kitaev candidates \cite{liu2018pseudospin,sano2018kitaev}. In these materials, the exchange processes via $t_{2g}$-$e_g$ and $e_g$-$e_g$ channels, which are absent in 4$d$ and 5$d$ candidates, suppress Heisenberg interactions while enhancing Kitaev interactions. However, similar to 4$d$ and 5$d$ Kitaev QSL candidates, all 3$d^7$ candidates exhibit antiferromagnetic order below Neel temperatures due to unavoidable non-Kitaev interactions \cite{doi:10.1126/sciadv.aay6953, PhysRevB.94.214416, PhysRevB.95.094424,VICIU20071060}. Therefore, it is still necessary to employ the second strategy to achieve a possible QSL state. 

Among the candidates, Na$_3$Co$_2$SbO$_6$ (NCSO) is considered hosting significant Kitaev interaction \cite{PhysRevLett.125.047201,Kim_2022,PhysRevB.107.054411,WONG201618,PhysRevMaterials.3.074405,C9NJ03627J}, although contradictory ferromagnetic ($K$ = -9 meV \cite{NSCO_neutron_Kitaev_parameter} and antiferromagnetic ($K$ = +3.6 meV \cite{Kim_3d}) estimates have been reported. At zero magnetic field, NCSO shows an antiferromagnetic order below $T_N$=6.6 K. Its exact magnetic structure is still a topic of debate, with proposed models including zigzag \cite{WONG201618, PhysRevMaterials.3.074405} and double-$\mathbf{q}$ configurations \cite{Gu_NCSO_double_Q}. The observed suppression of ordered magnetic moments \cite{PhysRevMaterials.3.074405} and theoretical analyses \cite{PhysRevLett.125.047201, QSL_XXZ_model, XXZ_model1} suggest that competing interactions may place NCSO near a QSL phase, where weak external perturbations could overcome the ordering tendency \cite{PhysRevLett.125.047201}. Indeed, magnetic susceptibility measurement reveals that an in-plane magnetic field can suppress the magnetic order\cite{li2022giant}, driving NCSO to a potential QSL state before transforming into a polarized state\cite{PhysRevB.109.054411}. 

 \begin{figure*}  [htpt]  
\includegraphics[clip, width=0.8\textwidth]{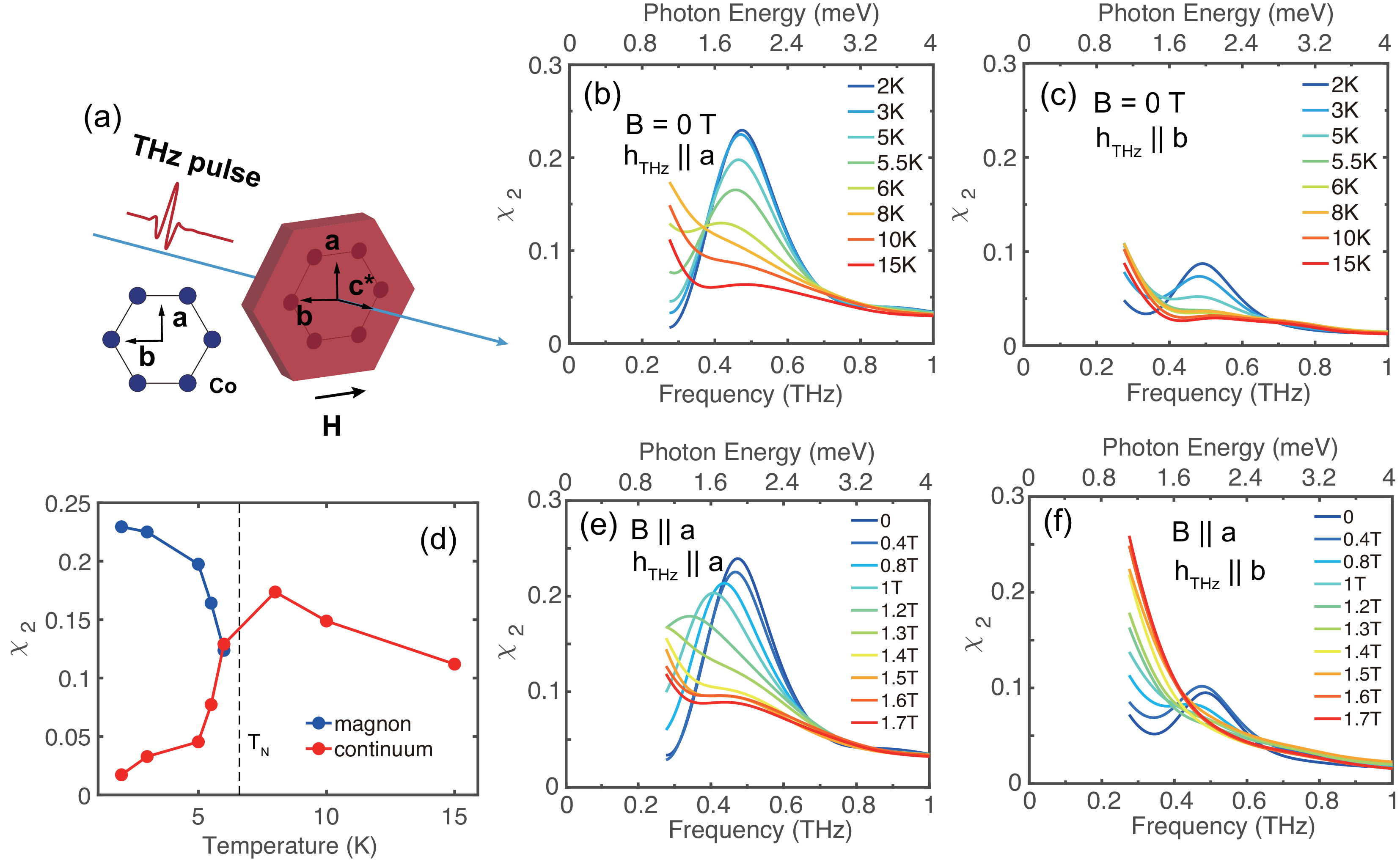}\\[1pt]
\caption{(a) Schematic of the THz transmission measurement under an external magnetic field. (b) (c) Imaginary part of magnetic susceptibility $\chi_2$($\omega$) at selected temperatures, with $\bf{h} _\mathbf{THz} \parallel \bf{a}$ and $\bf{h} _\mathbf{THz} \parallel \bf{b}$. (d) The temperature dependence of $\chi_2$ at 0.25 THz and 0.47 THz for $\bf{h} _\mathbf{THz} \parallel  \bf{a}$ geometry, indicates the intensity of the continuum and magnon, respectively. (e) (f) The magnetic field evolutions of $\chi_2$($\omega$) at 2 K, with $\bf{h} _\mathbf{THz}  \parallel \bf{a}$ and $\bf{h} _\mathbf{THz} \parallel \bf{b}$. The magnetic field ($\bf{B}$) is along the $\bf{a}$ axis.
\label{fig:1}}
\end{figure*}

While not a firm proof\cite{NaBaCoPO_neutron}, observation of a spin excitation continuum is often considered a strong indication of a QSL state. Given the ambiguous role of disorder in generating spectral continua \cite{Disorder_QSL,continuum_neutron_Cu}, spectroscopic studies on a structurally pristine single crystal are imperative to elucidate the origin of the continuum. NCSO is promising for this purpose: optimized synthesis methods yield twin-free single crystals without multiple monoclinic domains \cite{li2022giant}. Furthermore, these crystals exhibit in-plane C$_2$ symmetry, signaling an unusual anisotropy within the honeycomb plane\cite{li2022giant, Gu_NCSO_double_Q}.  Spectroscopic studies on such twin-free crystals under magnetic fields are essential to examine the nature of the field-induced disorder state and refine potential spin Hamiltonians for NCSO.

In this work, we performed terahertz (THz) time-domain spectroscopy on a twin-free single crystal of NCSO, probing spin excitations at the Brillouin zone center ($\mathbf{q} = 0$). As shown in Figs~\ref{fig:1}(a), the THz wave propagates along the $\mathbf{c^*}$ axis (perpendicular to the honeycomb plane). We observed a continuum-like spin excitation above $T_N$ and a collective spin excitation (magnon) at 0.5 THz (2 meV) below $T_N$. Upon applying an in-plane magnetic field within an intermediate range, the distinct magnon peak transforms into a continuum spectrum, which may indicate the emergence of fractionalized excitations. The low critical field and broad intermediate phase highlight NCSO as a promising platform for exploring the magnetic continuum. Interestingly,  the field evolutions of the spin excitations along $\bf{b}$ and $\bf{a}$ axes are rather distinct across different phases. These results highlight NCSO as an intriguing system with the potential to realize QSL and quantum spin models with an anisotropic structure.

The low-energy dynamics of NCSO are dominated by magnetic interactions, as its electronic degrees of freedom are frozen at low temperatures due to a large band gap. We first analyze the temperature dependence of the imaginary part of the dynamic spin susceptibility, $\chi_2(\omega)$, extracted from transmission data in the $\mathbf{h}_\mathrm{THz} \parallel \mathbf{a}$ geometry without external field (see Supplementary for details\cite{supplemental_material}). Above $T_N$, a low-energy continuum extending up to 0.8~THz (3.2~meV) emerges, with increasing spectral weight upon cooling, as shown in Fig.~\ref{fig:1}(b)(d). Similar continua have been observed in $\alpha$-RuCl$_3$ and BaCo$_2$(AsO$_4$)$_2$~\cite{banerjee2017neutron,Banerjee2018,zhang2023magnetic}. In the context of Kitaev QSL, such continua reflect fractionalized excitations below the spin-correlation temperature $T_K$, representing a hallmark of dominant Kitaev interactions~\cite{Knolle_theory_kitaev}. Other possible interpretations of the continuum will be discussed later.

Below $T_N \sim 6.6$ K, the spectral weight of continuum excitation collapses abruptly, and a narrow magnon peak emerges at 0.47 THz. At 6 K (slightly below $T_N$), the magnon and continuum excitation coexist, as shown in Fig.~\ref{fig:1} (b). Figure~\ref{fig:1} (d) shows the temperature dependence of $\chi _2$ at 0.25 THz and 0.47 THz, as indicators of the continuum and magnon, respectively. At 2 K, the magnon peak dominates, and the continuum excitation is negligible. We integrate the spectra weight of $\chi _2 (\omega)$ with $\chi _0$=$ \int_{0}^{\infty} \frac{\chi _2 (\omega)}{\omega} d \omega$ to estimate the contribution from the magnon to the $dc$ magnetic susceptibility\cite{PhysRevLett.119.227201}. Upon integrating the spectral weight up to 1 THz, the obtained value is $\sim$ 0.08 in SI units, comparable to that from $dc$ magnetometry (0.34 emu/mole Co/Oe $\approx$ 0.08 SI), indicating a full transfer of spectral weight into magnon excitations. This contrasts with the residual low-energy continuum observed in BaCo$_2$(AsO$_4$)$_2$ \cite{zhang2023magnetic}. As mentioned above, NCSO exhibits strong THz response anisotropy between the ${\bf a}$ and ${\bf b}$ axes. This anisotropy is already apparent in the zero-field spectra, as shown in Figs.~\ref{fig:1}(b) and (c). The magnon intensity for ${\bf h}_{\bf THz} \parallel {\bf a}$ is approximately twice that for ${\bf h}_{\bf THz} \parallel {\bf b}$. The THz magnetic field component excites spin waves (magnons) through the Zeeman interaction, preferentially coupling to magnons with moments perpendicular to ${\bf h}_{\bf THz}$. Neutron diffraction studies have suggested a double-${\bf q}$ magnetic structure or two equivalent zigzag domains in NCSO \cite{Gu_NCSO_double_Q}, in contrast to the three zigzag domains reported in RuCl$_3$ \cite{RuCl3_domains}, due to the strong in-plane anisotropy. The observed $\chi_2(\omega)$ intensity difference may directly reflect the absence of the third zigzag component in NCSO.

The long-range magnetic order in NCSO can be suppressed by an in-plane magnetic field \cite{li2022giant}. We begin by applying a magnetic field along the $\bf{a}$ axis and tracking the excitations at 2 K. As shown in Figs.~\ref{fig:1}(e,f), the magnon peak softens with increasing field and vanishes near the critical field. In the intermediate range (1.3–1.7~T), the spectrum is dominated by a low-energy continuum. This observation is qualitatively similar to the result in BaCo$_2$(AsO$_4$)$_2$ \cite{zhang2023magnetic}. However, a key difference emerges: In BaCo$_2$(AsO$_4$)$_2$, spins polarize immediately upon suppression of the antiferromagnetic order, and the continuum appears only in a narrow range ($\sim$ 0.5 T), making it difficult to determine whether it represents a true intermediate phase or merely a transition boundary between magnetic orders \cite{Shi2021}. In contrast, NCSO exhibits the continuum across a wide field range (1.3 - 1.7 T), demonstrating the robustness of this field-induced intermediate phase. 

\begin{figure}[htpt]
\includegraphics[clip, width=0.5\textwidth]{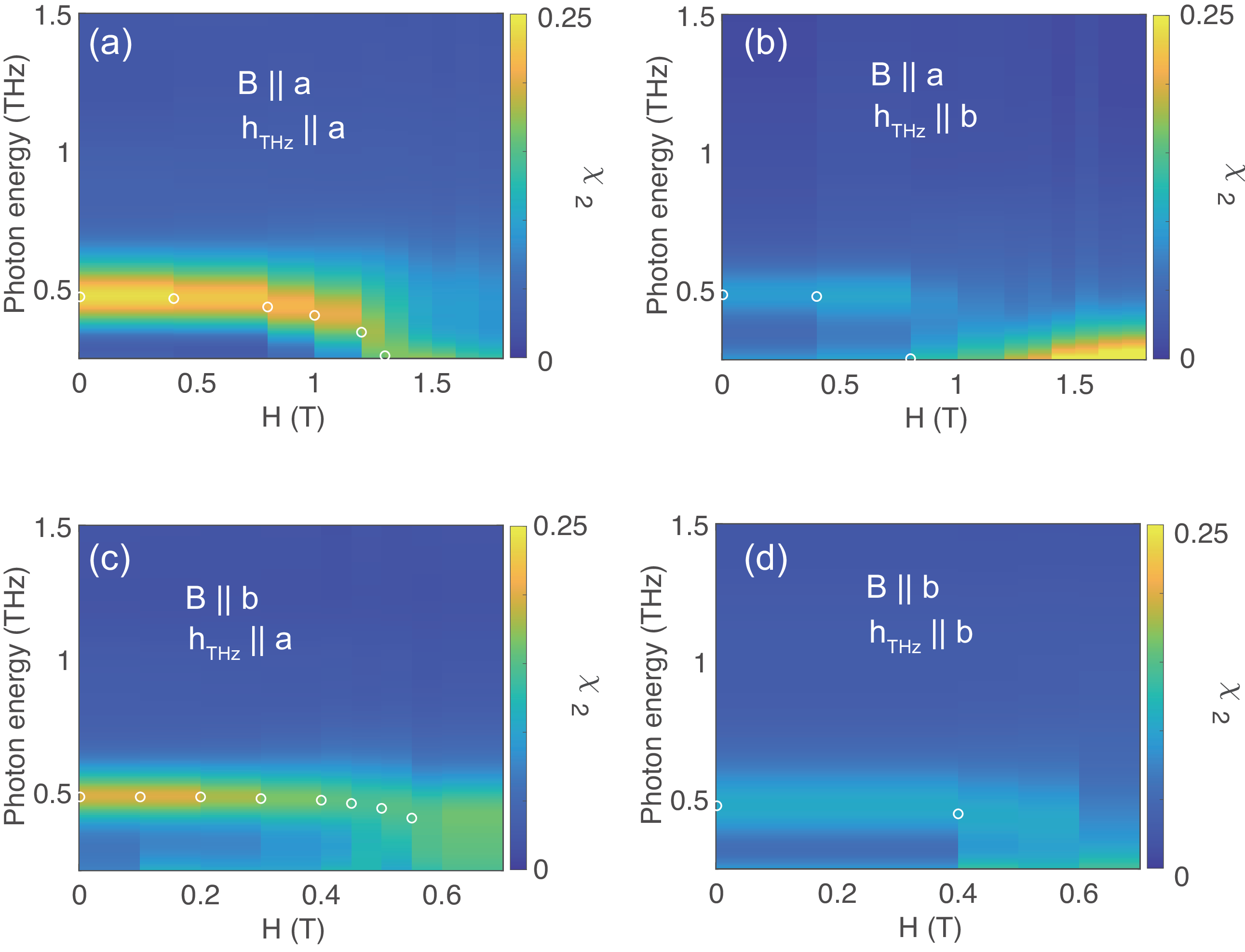}\\[1pt] 
\caption{Temperatures and fields evolution of the magnetic susceptibility $\chi_2$($\omega$). The white circles indicate the peak position of the modes. (a) Imaginary part of magnetic susceptibility $\chi_2$($\omega$) at low fields with $\mathbf{B}\parallel \mathbf{a} $ and $\bf{h} _\mathbf{THz}  \parallel \bf{a}$. (b) The magnetic field evolution of $\chi_2$($\omega$) at low fields with $\mathbf{B}\parallel \bf{a} $ and $\bf{h} _\mathbf{THz}  \parallel \bf{b}$. (c) The magnetic field evolution of $\chi_2$($\omega$) at low fields with $\mathbf{B}\parallel \bf{b}$ and $\bf{h} _\mathbf{THz}  \parallel \bf{a}$. (d) The magnetic field evolution of $\chi_2$($\omega$) at low fields with $\mathbf{B}\parallel \bf{b} $ and $\bf{h} _\mathbf{THz}  \parallel \bf{b}$. 
\label{fig:2}}
\end{figure}

\begin{figure*}[htpt]
\includegraphics[clip, width=1\textwidth]{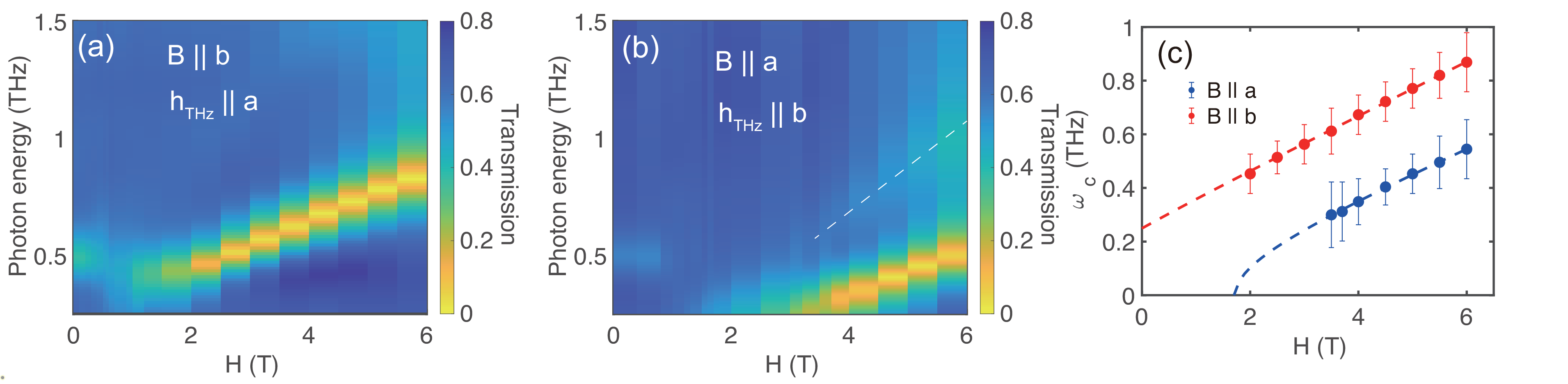}
\caption{The terahertz transmission spectrum with $\bf{h} _\mathbf{THz} \perp \bf{B}$ configurations: (a) and (b) are $\bf{B}$ along $\bf{b}$ and $\bf{a}$ axis, respectively. The white dashed line in (b) is indicative of the two-magnon excitations. (c) The frequency evolution of the spin waves in the spin-polarized phase. The error bars indicate the FWHM of the spin wave peaks. The dashed lines are the fitting curves with linear spin-wave theory described by equations \eqref{equ:epsilon_a}\eqref{equ:epsilon_b}. 
\label{fig:3}}
\end{figure*}

To further investigate the in-plane anisotropy, we compare spin excitations for ${\bf B} \parallel {\bf a}$ and ${\bf B} \parallel {\bf b}$ (Fig.~\ref{fig:2}). The lower critical fields are 1.3 T for ${\bf B} \parallel {\bf a}$ and 0.5 T for ${\bf B} \parallel {\bf b}$. For ${\bf B} \parallel {\bf a}$, the magnon softens near 1 T and vanishes above 1.3 T, replaced by a continuum. For ${\bf B} \parallel {\bf b}$, the magnon rapidly diminishes at 0.5 T without observable softening. These polarization- and field-dependent magnon measurements provide insights into the spin orientation. A simple zigzag structure with spins aligned along ${\bf b}$ \cite{PhysRevMaterials.3.074405} would predict that magnons are only observable for ${\bf h}_{\bf THz} \parallel {\bf a}$, contradicting our observations along both axes. The data can be reconciled by considering two equivalent zigzag domains tilted from ${\bf b}$ or a vector sum of them, leading to a double-$q$ \cite{Gu_NCSO_double_Q}. Notably, unlike RuCl$_3$ \cite{shi2018field, PhysRevB.98.094425}, we observe no magnon intensity enhancement in ${\bf B} \perp {\bf h}_{\bf THz}$ configurations, suggesting no domain repopulation occurs. While our results have no obvious contradiction with a double-$q$ structure, they don't definitively exclude the two-domain zigzag scenario. Future THz polarimetry with field training \cite{Gu_NCSO_double_Q} could help resolve the precise magnetic structure.


As the magnetic field is further increased, the system transitions into a spin-polarized state. The magnetic excitations in the polarized state are characterized by well-defined spin waves. Figures~\ref{fig:3}(a) and (b) show the terahertz transmission at 2 K for $\mathbf{B} \parallel \mathbf{b}$ and $\mathbf{B} \parallel \mathbf{a}$ configurations, respectively, with $\mathbf{h}_{\mathbf{THz}}$ perpendicular to the applied magnetic field. The evolution of one-magnon excitations is clearly identified by the light-yellow regions. Concurrently, broad two-magnon excitations are observed at twice the frequency of the one-magnon (see dashed-line indicators in Fig.~\ref{fig:3}(b). The one-magnon modes vanish in $\mathbf{B} \parallel \mathbf{h}_{\mathbf{THz}}$, leaving broad spectral features of the longitudinal two-magnon responses. See Supplemental Material (SM)\cite{supplemental_material} for full results. The observation of the two-magnon excitations in both configurations suggests the high-field phase has not been fully polarized.

Field-dependent mode frequencies extracted from peak positions are plotted in Fig.~\ref{fig:3}(c). The evolution of magnons under magnetic fields along different axes exhibits distinct behaviors, reflecting the anisotropic $g$-factor and magnetic interactions. While the magnetic model of NCSO remains undetermined, the magnetic interactions can be parameterized through Curie-Weiss temperatures. To describe the anisotropic evolution of these modes, we develop a phenomenological model incorporating anisotropic $g$-factors and Curie-Weiss temperatures\cite{supplemental_material}:

\begin{equation}
\label{equ:epsilon_a}
\epsilon_{a,+} = \sqrt{ 
\left( \mu_B B g_{a} + 2k_B\Theta_{a} - 2k_B\Theta_{b} \right)
\left( \mu_B B g_{a} + 2k_B\Theta_{a} - 2k_B\Theta_{c^*} \right) 
}
\end{equation}

\begin{equation}
\label{equ:epsilon_b}
\epsilon_{b,+} = \sqrt{ 
\left( \mu_B B g_{b} + 2k_B\Theta_{b} - 2k_B\Theta_{a} \right)
\left( \mu_B B g_{b} + 2k_B\Theta_{b} - 2k_B\Theta_{c^*} \right) 
}
\end{equation}

where $B$ is the magnetic field, $\mu_B$ is the Bohr magneton, $g_a$ and $g_b$ are anisotropic $g$-factors, $k_B$ is Boltzmann's constant, and $\Theta_a$, $\Theta_b$, $\Theta_{c^*}$ represent Curie-Weiss temperatures along the respective crystallographic axes.

\setlength{\tabcolsep}{10pt}
\renewcommand{\arraystretch}{1.5}
\begin{table}[h!]
\caption{\label{tab:table1}%
Fitting results of Equations~\eqref{equ:epsilon_a} and \eqref{equ:epsilon_b}.}
\begin{ruledtabular}
\begin{tabular}{cc}

Parameter & Value \\
\hline
$g_{a}$ & 6.31 \\
$g_{b}$ & 7.14 \\
$\Theta_{a} - \Theta_{b}$ & $-3.66$\,K \\
$\Theta_{b} - \Theta_{c^*}$ & $9.78$\,K \\

\end{tabular}
\end{ruledtabular}
\end{table}

The dashed lines in Fig.~\ref{fig:3}(c) represent the fitting curves, with corresponding parameters summarized in Table~\ref{tab:table1}. The extracted $g$-factors exhibit slight anisotropy, yielding $g = 6.31$ for $\mathbf{B} \parallel \mathbf{a}$ and $g = 7.14$ for $\mathbf{B} \parallel \mathbf{b}$. Our analysis establishes a direct correlation between the anisotropic Curie-Weiss temperatures and the field-dependent magnon energy evolution. These interaction parameters can be further constrained within specific theoretical frameworks, as discussed in SM\cite{supplemental_material}.

Comparing the observed continuum with other Kitaev candidates is instrumental in elucidating the nature of the excitations. The continuum in NCSO closely resembles that in BaCo$_2$(AsO$_4$)$_2$ \cite{zhang2023magnetic}, in terms of profile, energy scale, and temperature dependence, despite different magnetic ground states. This similarity suggests the continuum is dominated by the Kitaev or nearest-neighbor (J$_1$) interaction and is insensitive to the underlying magnetic structure. Motivated by their low-energy similarity, we examined the out-of-plane magnetic field response up to 15 T. Unlike BaCo$_2$(AsO$_4$)$_2$, where a disordered state emerges above 4.5 T \cite{zhang2023magnetic}, NCSO retains long-range order up to 10 T \cite{supplemental_material}, with $T_N$ only gradually suppressed. This contrast may be caused by the different magnetic structures. 
In comparison, RuCl$_3$ and Na$_2$Co$_2$TeO$_6$ host a broad continuum peaked at finite frequencies ~\cite{PhysRevLett.114.147201, PhysRevLett.119.227201, PhysRevLett.119.227202, shi2018field, PhysRevB.101.140410, PhysRevB.108.L140406, PhysRevLett.131.076701}. For the RuCl$_3$, spectral weight analysis suggests that the higher frequency continuum must come from electric dipole coupling since its spectral weight is much larger than a conventional magnetic dipole coupling \cite{shi2018field, PhysRevB.98.094425}. Notably, both BaCo$_2$(AsO$_4$)$_2$ and NCSO are free of stacking faults and twin domains, unlike RuCl$_3$ and Na$_2$Co$_2$TeO$_6$, where structural disorder may obfuscate the exploration of the intrinsic continuum excitations. 

Finally, we discuss the possible interpretations of the observed continuum. While the continuum may arise from Kitaev-type spin fractionalization, an alternative interpretation is based on the XXZ-J$_1$-J$_3$ model, where low-energy spinon excitations originating from a Dirac spin liquid state persist within a narrow region of the J$_3$/J$_1$ parameter space, leading to a continuum feature \cite{XXZ_model1}. Parton mean-field theory predicts a pronounced enhancement of the low-energy THz response as an asymmetric peak at $\omega/J_1 \sim 0.1$ \cite{XXZ_model1}. We note that there is still room for
the existence of such a peak in the low-energy part of the $\chi_2(\omega)$ not covered by our measurement window. Applying the spectral weight analysis to $\chi_2(\omega)$ yields  $\chi_0 \backsim $0.1 (SI unit) at 7 K, which is lower than the measured value of static susceptibility 0.17 in Systeme International units (0.7 emu/mole Co/Oe). The sum rule analysis suggests the existence of a small fraction of continuum excitations below the low limit of our measured range. Future microwave spectroscopy could search for the predicted low-energy peak and further constrain the spin Hamiltonian. Furthermore, the existence of a QSL state in the XXZ-$J_1$-$J_3$ model remains contentious, as recent density-matrix renormalization group studies find no evidence for such a phase\cite{DMRG_J1_J3}. Nevertheless, significant quantum fluctuations persist within the magnetically ordered regime, likely originating from Kitaev or frustrated $J_3$ interactions\cite{NCSO_residual_fluctuations}. The microscopic origin of this continuum and the corresponding spin structure warrant further experimental and theoretical investigations.

In this study, we systematically investigated low-energy magnetic excitations in NCSO using time-domain terahertz spectroscopy. Signatures of the field-induced disorder state and magnetic continuum excitations are observed. 
Notably, rotating the magnetic field from \textbf{a} axis to \textbf{b} axis reveals strong anisotropic evolution of the spin wave excitations. Our results establish NCSO as an exceptional platform for studying field-induced magnetic continua and potential QSL states. The strong in-plane anisotropy provides unique opportunities for investigating anisotropic quantum spin models.

\begin{acknowledgments}
We are grateful for discussions with Yuan Wan. This work was supported by National Natural Science Foundation of China (Grant Nos.12488201,12250008), the National Key Research and Development Program of China (Grant No. 2022YFA1403901).
\end{acknowledgments}

\bibliographystyle{apsrev4-2}

\bibliography{ref}

\begin{thebibliography}{71}%
\makeatletter
\providecommand \@ifxundefined [1]{%
 \@ifx{#1\undefined}
}%
\providecommand \@ifnum [1]{%
 \ifnum #1\expandafter \@firstoftwo
 \else \expandafter \@secondoftwo
 \fi
}%
\providecommand \@ifx [1]{%
 \ifx #1\expandafter \@firstoftwo
 \else \expandafter \@secondoftwo
 \fi
}%
\providecommand \natexlab [1]{#1}%
\providecommand \enquote  [1]{``#1''}%
\providecommand \bibnamefont  [1]{#1}%
\providecommand \bibfnamefont [1]{#1}%
\providecommand \citenamefont [1]{#1}%
\providecommand \href@noop [0]{\@secondoftwo}%
\providecommand \href [0]{\begingroup \@sanitize@url \@href}%
\providecommand \@href[1]{\@@startlink{#1}\@@href}%
\providecommand \@@href[1]{\endgroup#1\@@endlink}%
\providecommand \@sanitize@url [0]{\catcode `\\12\catcode `\$12\catcode `\&12\catcode `\#12\catcode `\^12\catcode `\_12\catcode `\%12\relax}%
\providecommand \@@startlink[1]{}%
\providecommand \@@endlink[0]{}%
\providecommand \url  [0]{\begingroup\@sanitize@url \@url }%
\providecommand \@url [1]{\endgroup\@href {#1}{\urlprefix }}%
\providecommand \urlprefix  [0]{URL }%
\providecommand \Eprint [0]{\href }%
\providecommand \doibase [0]{https://doi.org/}%
\providecommand \selectlanguage [0]{\@gobble}%
\providecommand \bibinfo  [0]{\@secondoftwo}%
\providecommand \bibfield  [0]{\@secondoftwo}%
\providecommand \translation [1]{[#1]}%
\providecommand \BibitemOpen [0]{}%
\providecommand \bibitemStop [0]{}%
\providecommand \bibitemNoStop [0]{.\EOS\space}%
\providecommand \EOS [0]{\spacefactor3000\relax}%
\providecommand \BibitemShut  [1]{\csname bibitem#1\endcsname}%
\let\auto@bib@innerbib\@empty
\bibitem [{\citenamefont {Anderson}(1973)}]{ANDERSON1973153}%
  \BibitemOpen
  \bibfield  {author} {\bibinfo {author} {\bibfnamefont {P.}~\bibnamefont {Anderson}},\ }\href {https://doi.org/https://doi.org/10.1016/0025-5408(73)90167-0} {\bibfield  {journal} {\bibinfo  {journal} {Materials Research Bulletin}\ }\textbf {\bibinfo {volume} {8}},\ \bibinfo {pages} {153} (\bibinfo {year} {1973})}\BibitemShut {NoStop}%
\bibitem [{\citenamefont {Balents}(2010)}]{balents2010spin}%
  \BibitemOpen
  \bibfield  {author} {\bibinfo {author} {\bibfnamefont {L.}~\bibnamefont {Balents}},\ }\href {https://doi.org/10.1038/nature08917} {\bibfield  {journal} {\bibinfo  {journal} {nature}\ }\textbf {\bibinfo {volume} {464}},\ \bibinfo {pages} {199} (\bibinfo {year} {2010})}\BibitemShut {NoStop}%
\bibitem [{\citenamefont {Knolle}\ and\ \citenamefont {Moessner}(2019)}]{knolle2019field}%
  \BibitemOpen
  \bibfield  {author} {\bibinfo {author} {\bibfnamefont {J.}~\bibnamefont {Knolle}}\ and\ \bibinfo {author} {\bibfnamefont {R.}~\bibnamefont {Moessner}},\ }\href {https://doi.org/10.1146/annurev-conmatphys-031218-013401} {\bibfield  {journal} {\bibinfo  {journal} {Annual Review of Condensed Matter Physics}\ }\textbf {\bibinfo {volume} {10}},\ \bibinfo {pages} {451} (\bibinfo {year} {2019})}\BibitemShut {NoStop}%
\bibitem [{\citenamefont {Takagi}\ \emph {et~al.}(2019{\natexlab{a}})\citenamefont {Takagi}, \citenamefont {Takayama}, \citenamefont {Jackeli}, \citenamefont {Khaliullin},\ and\ \citenamefont {Nagler}}]{takagi2019concept}%
  \BibitemOpen
  \bibfield  {author} {\bibinfo {author} {\bibfnamefont {H.}~\bibnamefont {Takagi}}, \bibinfo {author} {\bibfnamefont {T.}~\bibnamefont {Takayama}}, \bibinfo {author} {\bibfnamefont {G.}~\bibnamefont {Jackeli}}, \bibinfo {author} {\bibfnamefont {G.}~\bibnamefont {Khaliullin}},\ and\ \bibinfo {author} {\bibfnamefont {S.~E.}\ \bibnamefont {Nagler}},\ }\href {https://doi.org/10.1038/s42254-019-0038-2} {\bibfield  {journal} {\bibinfo  {journal} {Nature Reviews Physics}\ }\textbf {\bibinfo {volume} {1}},\ \bibinfo {pages} {264} (\bibinfo {year} {2019}{\natexlab{a}})}\BibitemShut {NoStop}%
\bibitem [{\citenamefont {Freedman}\ \emph {et~al.}(2003)\citenamefont {Freedman}, \citenamefont {Kitaev}, \citenamefont {Larsen},\ and\ \citenamefont {Wang}}]{freedman2003topological}%
  \BibitemOpen
  \bibfield  {author} {\bibinfo {author} {\bibfnamefont {M.}~\bibnamefont {Freedman}}, \bibinfo {author} {\bibfnamefont {A.}~\bibnamefont {Kitaev}}, \bibinfo {author} {\bibfnamefont {M.}~\bibnamefont {Larsen}},\ and\ \bibinfo {author} {\bibfnamefont {Z.}~\bibnamefont {Wang}},\ }\href {https://www.ams.org/journals/bull/2003-40-01/S0273-0979-02-00964-3/} {\bibfield  {journal} {\bibinfo  {journal} {Bulletin of the American Mathematical Society}\ }\textbf {\bibinfo {volume} {40}},\ \bibinfo {pages} {31} (\bibinfo {year} {2003})}\BibitemShut {NoStop}%
\bibitem [{\citenamefont {Motome}\ and\ \citenamefont {Nasu}(2020)}]{motome2020hunting}%
  \BibitemOpen
  \bibfield  {author} {\bibinfo {author} {\bibfnamefont {Y.}~\bibnamefont {Motome}}\ and\ \bibinfo {author} {\bibfnamefont {J.}~\bibnamefont {Nasu}},\ }\href {https://doi.org/10.7566/JPSJ.89.012002} {\bibfield  {journal} {\bibinfo  {journal} {Journal of the Physical Society of Japan}\ }\textbf {\bibinfo {volume} {89}},\ \bibinfo {pages} {012002} (\bibinfo {year} {2020})}\BibitemShut {NoStop}%
\bibitem [{\citenamefont {Kitaev}(2006)}]{KITAEV20062}%
  \BibitemOpen
  \bibfield  {author} {\bibinfo {author} {\bibfnamefont {A.}~\bibnamefont {Kitaev}},\ }\href {https://doi.org/https://doi.org/10.1016/j.aop.2005.10.005} {\bibfield  {journal} {\bibinfo  {journal} {Annals of Physics}\ }\textbf {\bibinfo {volume} {321}},\ \bibinfo {pages} {2} (\bibinfo {year} {2006})},\ \bibinfo {note} {january Special Issue}\BibitemShut {NoStop}%
\bibitem [{\citenamefont {Banerjee}\ \emph {et~al.}(2017)\citenamefont {Banerjee}, \citenamefont {Yan}, \citenamefont {Knolle}, \citenamefont {Bridges}, \citenamefont {Stone}, \citenamefont {Lumsden}, \citenamefont {Mandrus}, \citenamefont {Tennant}, \citenamefont {Moessner},\ and\ \citenamefont {Nagler}}]{banerjee2017neutron}%
  \BibitemOpen
  \bibfield  {author} {\bibinfo {author} {\bibfnamefont {A.}~\bibnamefont {Banerjee}}, \bibinfo {author} {\bibfnamefont {J.}~\bibnamefont {Yan}}, \bibinfo {author} {\bibfnamefont {J.}~\bibnamefont {Knolle}}, \bibinfo {author} {\bibfnamefont {C.~A.}\ \bibnamefont {Bridges}}, \bibinfo {author} {\bibfnamefont {M.~B.}\ \bibnamefont {Stone}}, \bibinfo {author} {\bibfnamefont {M.~D.}\ \bibnamefont {Lumsden}}, \bibinfo {author} {\bibfnamefont {D.~G.}\ \bibnamefont {Mandrus}}, \bibinfo {author} {\bibfnamefont {D.~A.}\ \bibnamefont {Tennant}}, \bibinfo {author} {\bibfnamefont {R.}~\bibnamefont {Moessner}},\ and\ \bibinfo {author} {\bibfnamefont {S.~E.}\ \bibnamefont {Nagler}},\ }\href {https://doi.org/10.1126/science.aah6015} {\bibfield  {journal} {\bibinfo  {journal} {Science}\ }\textbf {\bibinfo {volume} {356}},\ \bibinfo {pages} {1055} (\bibinfo {year} {2017})},\ \Eprint {https://arxiv.org/abs/https://www.science.org/doi/pdf/10.1126/science.aah6015} {https://www.science.org/doi/pdf/10.1126/science.aah6015}
  \BibitemShut {NoStop}%
\bibitem [{\citenamefont {Hermanns}\ \emph {et~al.}(2018)\citenamefont {Hermanns}, \citenamefont {Kimchi},\ and\ \citenamefont {Knolle}}]{Hermanns2018}%
  \BibitemOpen
  \bibfield  {author} {\bibinfo {author} {\bibfnamefont {M.}~\bibnamefont {Hermanns}}, \bibinfo {author} {\bibfnamefont {I.}~\bibnamefont {Kimchi}},\ and\ \bibinfo {author} {\bibfnamefont {J.}~\bibnamefont {Knolle}},\ }\href {https://doi.org/https://doi.org/10.1146/annurev-conmatphys-033117-053934} {\bibfield  {journal} {\bibinfo  {journal} {Annual Review of Condensed Matter Physics}\ }\textbf {\bibinfo {volume} {9}},\ \bibinfo {pages} {17} (\bibinfo {year} {2018})}\BibitemShut {NoStop}%
\bibitem [{\citenamefont {Takagi}\ \emph {et~al.}(2019{\natexlab{b}})\citenamefont {Takagi}, \citenamefont {Takayama}, \citenamefont {Jackeli}, \citenamefont {Khaliullin},\ and\ \citenamefont {Nagler}}]{Takagi2019}%
  \BibitemOpen
  \bibfield  {author} {\bibinfo {author} {\bibfnamefont {H.}~\bibnamefont {Takagi}}, \bibinfo {author} {\bibfnamefont {T.}~\bibnamefont {Takayama}}, \bibinfo {author} {\bibfnamefont {G.}~\bibnamefont {Jackeli}}, \bibinfo {author} {\bibfnamefont {G.}~\bibnamefont {Khaliullin}},\ and\ \bibinfo {author} {\bibfnamefont {S.~E.}\ \bibnamefont {Nagler}},\ }\href {https://doi.org/10.1038/s42254-019-0038-2} {\bibfield  {journal} {\bibinfo  {journal} {Nature Reviews Physics}\ }\textbf {\bibinfo {volume} {1}},\ \bibinfo {pages} {264} (\bibinfo {year} {2019}{\natexlab{b}})}\BibitemShut {NoStop}%
\bibitem [{\citenamefont {Nasu}\ \emph {et~al.}(2017)\citenamefont {Nasu}, \citenamefont {Yoshitake},\ and\ \citenamefont {Motome}}]{PhysRevLett.119.127204}%
  \BibitemOpen
  \bibfield  {author} {\bibinfo {author} {\bibfnamefont {J.}~\bibnamefont {Nasu}}, \bibinfo {author} {\bibfnamefont {J.}~\bibnamefont {Yoshitake}},\ and\ \bibinfo {author} {\bibfnamefont {Y.}~\bibnamefont {Motome}},\ }\href {https://doi.org/10.1103/PhysRevLett.119.127204} {\bibfield  {journal} {\bibinfo  {journal} {Phys. Rev. Lett.}\ }\textbf {\bibinfo {volume} {119}},\ \bibinfo {pages} {127204} (\bibinfo {year} {2017})}\BibitemShut {NoStop}%
\bibitem [{\citenamefont {Banerjee}\ \emph {et~al.}(2018)\citenamefont {Banerjee}, \citenamefont {Lampen-Kelley}, \citenamefont {Knolle}, \citenamefont {Balz}, \citenamefont {Aczel}, \citenamefont {Winn}, \citenamefont {Liu}, \citenamefont {Pajerowski}, \citenamefont {Yan}, \citenamefont {Bridges}, \citenamefont {Savici}, \citenamefont {Chakoumakos}, \citenamefont {Lumsden}, \citenamefont {Tennant}, \citenamefont {Moessner}, \citenamefont {Mandrus},\ and\ \citenamefont {Nagler}}]{Banerjee2018}%
  \BibitemOpen
  \bibfield  {author} {\bibinfo {author} {\bibfnamefont {A.}~\bibnamefont {Banerjee}}, \bibinfo {author} {\bibfnamefont {P.}~\bibnamefont {Lampen-Kelley}}, \bibinfo {author} {\bibfnamefont {J.}~\bibnamefont {Knolle}}, \bibinfo {author} {\bibfnamefont {C.}~\bibnamefont {Balz}}, \bibinfo {author} {\bibfnamefont {A.~A.}\ \bibnamefont {Aczel}}, \bibinfo {author} {\bibfnamefont {B.}~\bibnamefont {Winn}}, \bibinfo {author} {\bibfnamefont {Y.}~\bibnamefont {Liu}}, \bibinfo {author} {\bibfnamefont {D.}~\bibnamefont {Pajerowski}}, \bibinfo {author} {\bibfnamefont {J.}~\bibnamefont {Yan}}, \bibinfo {author} {\bibfnamefont {C.~A.}\ \bibnamefont {Bridges}}, \bibinfo {author} {\bibfnamefont {A.~T.}\ \bibnamefont {Savici}}, \bibinfo {author} {\bibfnamefont {B.~C.}\ \bibnamefont {Chakoumakos}}, \bibinfo {author} {\bibfnamefont {M.~D.}\ \bibnamefont {Lumsden}}, \bibinfo {author} {\bibfnamefont {D.~A.}\ \bibnamefont {Tennant}}, \bibinfo {author} {\bibfnamefont {R.}~\bibnamefont {Moessner}}, \bibinfo {author} {\bibfnamefont
  {D.~G.}\ \bibnamefont {Mandrus}},\ and\ \bibinfo {author} {\bibfnamefont {S.~E.}\ \bibnamefont {Nagler}},\ }\href {https://doi.org/10.1038/s41535-018-0079-2} {\bibfield  {journal} {\bibinfo  {journal} {npj Quantum Materials}\ }\textbf {\bibinfo {volume} {3}},\ \bibinfo {pages} {8} (\bibinfo {year} {2018})}\BibitemShut {NoStop}%
\bibitem [{\citenamefont {Knolle}\ \emph {et~al.}(2014{\natexlab{a}})\citenamefont {Knolle}, \citenamefont {Kovrizhin}, \citenamefont {Chalker},\ and\ \citenamefont {Moessner}}]{PhysRevLett.112.207203}%
  \BibitemOpen
  \bibfield  {author} {\bibinfo {author} {\bibfnamefont {J.}~\bibnamefont {Knolle}}, \bibinfo {author} {\bibfnamefont {D.~L.}\ \bibnamefont {Kovrizhin}}, \bibinfo {author} {\bibfnamefont {J.~T.}\ \bibnamefont {Chalker}},\ and\ \bibinfo {author} {\bibfnamefont {R.}~\bibnamefont {Moessner}},\ }\href {https://doi.org/10.1103/PhysRevLett.112.207203} {\bibfield  {journal} {\bibinfo  {journal} {Phys. Rev. Lett.}\ }\textbf {\bibinfo {volume} {112}},\ \bibinfo {pages} {207203} (\bibinfo {year} {2014}{\natexlab{a}})}\BibitemShut {NoStop}%
\bibitem [{\citenamefont {Winter}\ \emph {et~al.}(2017)\citenamefont {Winter}, \citenamefont {Riedl}, \citenamefont {Maksimov}, \citenamefont {Chernyshev}, \citenamefont {Honecker},\ and\ \citenamefont {Valentí}}]{Winter2017}%
  \BibitemOpen
  \bibfield  {author} {\bibinfo {author} {\bibfnamefont {S.~M.}\ \bibnamefont {Winter}}, \bibinfo {author} {\bibfnamefont {K.}~\bibnamefont {Riedl}}, \bibinfo {author} {\bibfnamefont {P.~A.}\ \bibnamefont {Maksimov}}, \bibinfo {author} {\bibfnamefont {A.~L.}\ \bibnamefont {Chernyshev}}, \bibinfo {author} {\bibfnamefont {A.}~\bibnamefont {Honecker}},\ and\ \bibinfo {author} {\bibfnamefont {R.}~\bibnamefont {Valentí}},\ }\href {https://doi.org/10.1038/s41467-017-01177-0} {\bibfield  {journal} {\bibinfo  {journal} {Nature Communications}\ }\textbf {\bibinfo {volume} {8}},\ \bibinfo {pages} {1152} (\bibinfo {year} {2017})}\BibitemShut {NoStop}%
\bibitem [{\citenamefont {Hickey}\ and\ \citenamefont {Trebst}(2019)}]{Hickey2019}%
  \BibitemOpen
  \bibfield  {author} {\bibinfo {author} {\bibfnamefont {C.}~\bibnamefont {Hickey}}\ and\ \bibinfo {author} {\bibfnamefont {S.}~\bibnamefont {Trebst}},\ }\href {https://doi.org/10.1038/s41467-019-08459-9} {\bibfield  {journal} {\bibinfo  {journal} {Nature Communications}\ }\textbf {\bibinfo {volume} {10}},\ \bibinfo {pages} {530} (\bibinfo {year} {2019})}\BibitemShut {NoStop}%
\bibitem [{\citenamefont {Do}\ \emph {et~al.}(2017)\citenamefont {Do}, \citenamefont {Park}, \citenamefont {Yoshitake}, \citenamefont {Nasu}, \citenamefont {Motome}, \citenamefont {Kwon}, \citenamefont {Adroja}, \citenamefont {Voneshen}, \citenamefont {Kim}, \citenamefont {Jang}, \citenamefont {Park}, \citenamefont {Choi},\ and\ \citenamefont {Ji}}]{Do2017}%
  \BibitemOpen
  \bibfield  {author} {\bibinfo {author} {\bibfnamefont {S.-H.}\ \bibnamefont {Do}}, \bibinfo {author} {\bibfnamefont {S.-Y.}\ \bibnamefont {Park}}, \bibinfo {author} {\bibfnamefont {J.}~\bibnamefont {Yoshitake}}, \bibinfo {author} {\bibfnamefont {J.}~\bibnamefont {Nasu}}, \bibinfo {author} {\bibfnamefont {Y.}~\bibnamefont {Motome}}, \bibinfo {author} {\bibfnamefont {Y.}~\bibnamefont {Kwon}}, \bibinfo {author} {\bibfnamefont {D.~T.}\ \bibnamefont {Adroja}}, \bibinfo {author} {\bibfnamefont {D.~J.}\ \bibnamefont {Voneshen}}, \bibinfo {author} {\bibfnamefont {K.}~\bibnamefont {Kim}}, \bibinfo {author} {\bibfnamefont {T.-H.}\ \bibnamefont {Jang}}, \bibinfo {author} {\bibfnamefont {J.-H.}\ \bibnamefont {Park}}, \bibinfo {author} {\bibfnamefont {K.-Y.}\ \bibnamefont {Choi}},\ and\ \bibinfo {author} {\bibfnamefont {S.}~\bibnamefont {Ji}},\ }\href {https://doi.org/10.1038/nphys4264} {\bibfield  {journal} {\bibinfo  {journal} {Nature Physics}\ }\textbf {\bibinfo {volume} {13}},\ \bibinfo {pages} {1079} (\bibinfo
  {year} {2017})}\BibitemShut {NoStop}%
\bibitem [{\citenamefont {Wulferding}\ \emph {et~al.}(2020)\citenamefont {Wulferding}, \citenamefont {Choi}, \citenamefont {Do}, \citenamefont {Lee}, \citenamefont {Lemmens}, \citenamefont {Faugeras}, \citenamefont {Gallais},\ and\ \citenamefont {Choi}}]{Wulferding2020}%
  \BibitemOpen
  \bibfield  {author} {\bibinfo {author} {\bibfnamefont {D.}~\bibnamefont {Wulferding}}, \bibinfo {author} {\bibfnamefont {Y.}~\bibnamefont {Choi}}, \bibinfo {author} {\bibfnamefont {S.-H.}\ \bibnamefont {Do}}, \bibinfo {author} {\bibfnamefont {C.~H.}\ \bibnamefont {Lee}}, \bibinfo {author} {\bibfnamefont {P.}~\bibnamefont {Lemmens}}, \bibinfo {author} {\bibfnamefont {C.}~\bibnamefont {Faugeras}}, \bibinfo {author} {\bibfnamefont {Y.}~\bibnamefont {Gallais}},\ and\ \bibinfo {author} {\bibfnamefont {K.-Y.}\ \bibnamefont {Choi}},\ }\href {https://doi.org/10.1038/s41467-020-15370-1} {\bibfield  {journal} {\bibinfo  {journal} {Nature Communications}\ }\textbf {\bibinfo {volume} {11}},\ \bibinfo {pages} {1603} (\bibinfo {year} {2020})}\BibitemShut {NoStop}%
\bibitem [{\citenamefont {Bruin}\ \emph {et~al.}(2022)\citenamefont {Bruin}, \citenamefont {Claus}, \citenamefont {Matsumoto}, \citenamefont {Kurita}, \citenamefont {Tanaka},\ and\ \citenamefont {Takagi}}]{Bruin2022}%
  \BibitemOpen
  \bibfield  {author} {\bibinfo {author} {\bibfnamefont {J.~A.~N.}\ \bibnamefont {Bruin}}, \bibinfo {author} {\bibfnamefont {R.~R.}\ \bibnamefont {Claus}}, \bibinfo {author} {\bibfnamefont {Y.}~\bibnamefont {Matsumoto}}, \bibinfo {author} {\bibfnamefont {N.}~\bibnamefont {Kurita}}, \bibinfo {author} {\bibfnamefont {H.}~\bibnamefont {Tanaka}},\ and\ \bibinfo {author} {\bibfnamefont {H.}~\bibnamefont {Takagi}},\ }\href {https://doi.org/10.1038/s41567-021-01501-y} {\bibfield  {journal} {\bibinfo  {journal} {Nature Physics}\ }\textbf {\bibinfo {volume} {18}},\ \bibinfo {pages} {401} (\bibinfo {year} {2022})}\BibitemShut {NoStop}%
\bibitem [{\citenamefont {Czajka}\ \emph {et~al.}(2023)\citenamefont {Czajka}, \citenamefont {Gao}, \citenamefont {Hirschberger}, \citenamefont {Lampen-Kelley}, \citenamefont {Banerjee}, \citenamefont {Quirk}, \citenamefont {Mandrus}, \citenamefont {Nagler},\ and\ \citenamefont {Ong}}]{Czajka2023}%
  \BibitemOpen
  \bibfield  {author} {\bibinfo {author} {\bibfnamefont {P.}~\bibnamefont {Czajka}}, \bibinfo {author} {\bibfnamefont {T.}~\bibnamefont {Gao}}, \bibinfo {author} {\bibfnamefont {M.}~\bibnamefont {Hirschberger}}, \bibinfo {author} {\bibfnamefont {P.}~\bibnamefont {Lampen-Kelley}}, \bibinfo {author} {\bibfnamefont {A.}~\bibnamefont {Banerjee}}, \bibinfo {author} {\bibfnamefont {N.}~\bibnamefont {Quirk}}, \bibinfo {author} {\bibfnamefont {D.~G.}\ \bibnamefont {Mandrus}}, \bibinfo {author} {\bibfnamefont {S.~E.}\ \bibnamefont {Nagler}},\ and\ \bibinfo {author} {\bibfnamefont {N.~P.}\ \bibnamefont {Ong}},\ }\href {https://doi.org/10.1038/s41563-022-01397-w} {\bibfield  {journal} {\bibinfo  {journal} {Nature Materials}\ }\textbf {\bibinfo {volume} {22}},\ \bibinfo {pages} {36} (\bibinfo {year} {2023})}\BibitemShut {NoStop}%
\bibitem [{\citenamefont {Wen}\ \emph {et~al.}(2019)\citenamefont {Wen}, \citenamefont {Yu}, \citenamefont {Li}, \citenamefont {Yu},\ and\ \citenamefont {Li}}]{Wen2019}%
  \BibitemOpen
  \bibfield  {author} {\bibinfo {author} {\bibfnamefont {J.}~\bibnamefont {Wen}}, \bibinfo {author} {\bibfnamefont {S.-L.}\ \bibnamefont {Yu}}, \bibinfo {author} {\bibfnamefont {S.}~\bibnamefont {Li}}, \bibinfo {author} {\bibfnamefont {W.}~\bibnamefont {Yu}},\ and\ \bibinfo {author} {\bibfnamefont {J.-X.}\ \bibnamefont {Li}},\ }\href {https://doi.org/10.1038/s41535-019-0151-6} {\bibfield  {journal} {\bibinfo  {journal} {npj Quantum Materials}\ }\textbf {\bibinfo {volume} {4}},\ \bibinfo {pages} {12} (\bibinfo {year} {2019})}\BibitemShut {NoStop}%
\bibitem [{\citenamefont {Witczak-Krempa}\ \emph {et~al.}(2014)\citenamefont {Witczak-Krempa}, \citenamefont {Chen}, \citenamefont {Kim},\ and\ \citenamefont {Balents}}]{witczak2014correlated}%
  \BibitemOpen
  \bibfield  {author} {\bibinfo {author} {\bibfnamefont {W.}~\bibnamefont {Witczak-Krempa}}, \bibinfo {author} {\bibfnamefont {G.}~\bibnamefont {Chen}}, \bibinfo {author} {\bibfnamefont {Y.~B.}\ \bibnamefont {Kim}},\ and\ \bibinfo {author} {\bibfnamefont {L.}~\bibnamefont {Balents}},\ }\href {https://doi.org/10.1146/annurev-conmatphys-020911-125138} {\bibfield  {journal} {\bibinfo  {journal} {Annual Review of Condensed Matter Physics}\ }\textbf {\bibinfo {volume} {5}},\ \bibinfo {pages} {57} (\bibinfo {year} {2014})},\ \Eprint {https://arxiv.org/abs/https://doi.org/10.1146/annurev-conmatphys-020911-125138} {https://doi.org/10.1146/annurev-conmatphys-020911-125138} \BibitemShut {NoStop}%
\bibitem [{\citenamefont {Plumb}\ \emph {et~al.}(2014)\citenamefont {Plumb}, \citenamefont {Clancy}, \citenamefont {Sandilands}, \citenamefont {Shankar}, \citenamefont {Hu}, \citenamefont {Burch}, \citenamefont {Kee},\ and\ \citenamefont {Kim}}]{PhysRevB.90.041112}%
  \BibitemOpen
  \bibfield  {author} {\bibinfo {author} {\bibfnamefont {K.~W.}\ \bibnamefont {Plumb}}, \bibinfo {author} {\bibfnamefont {J.~P.}\ \bibnamefont {Clancy}}, \bibinfo {author} {\bibfnamefont {L.~J.}\ \bibnamefont {Sandilands}}, \bibinfo {author} {\bibfnamefont {V.~V.}\ \bibnamefont {Shankar}}, \bibinfo {author} {\bibfnamefont {Y.~F.}\ \bibnamefont {Hu}}, \bibinfo {author} {\bibfnamefont {K.~S.}\ \bibnamefont {Burch}}, \bibinfo {author} {\bibfnamefont {H.-Y.}\ \bibnamefont {Kee}},\ and\ \bibinfo {author} {\bibfnamefont {Y.-J.}\ \bibnamefont {Kim}},\ }\href {https://doi.org/10.1103/PhysRevB.90.041112} {\bibfield  {journal} {\bibinfo  {journal} {Phys. Rev. B}\ }\textbf {\bibinfo {volume} {90}},\ \bibinfo {pages} {041112} (\bibinfo {year} {2014})}\BibitemShut {NoStop}%
\bibitem [{\citenamefont {Chaloupka}\ \emph {et~al.}(2010)\citenamefont {Chaloupka}, \citenamefont {Jackeli},\ and\ \citenamefont {Khaliullin}}]{PhysRevLett.105.027204}%
  \BibitemOpen
  \bibfield  {author} {\bibinfo {author} {\bibfnamefont {J.~c.~v.}\ \bibnamefont {Chaloupka}}, \bibinfo {author} {\bibfnamefont {G.}~\bibnamefont {Jackeli}},\ and\ \bibinfo {author} {\bibfnamefont {G.}~\bibnamefont {Khaliullin}},\ }\href {https://doi.org/10.1103/PhysRevLett.105.027204} {\bibfield  {journal} {\bibinfo  {journal} {Phys. Rev. Lett.}\ }\textbf {\bibinfo {volume} {105}},\ \bibinfo {pages} {027204} (\bibinfo {year} {2010})}\BibitemShut {NoStop}%
\bibitem [{\citenamefont {Kimchi}\ and\ \citenamefont {You}(2011)}]{PhysRevB.84.180407}%
  \BibitemOpen
  \bibfield  {author} {\bibinfo {author} {\bibfnamefont {I.}~\bibnamefont {Kimchi}}\ and\ \bibinfo {author} {\bibfnamefont {Y.-Z.}\ \bibnamefont {You}},\ }\href {https://doi.org/10.1103/PhysRevB.84.180407} {\bibfield  {journal} {\bibinfo  {journal} {Phys. Rev. B}\ }\textbf {\bibinfo {volume} {84}},\ \bibinfo {pages} {180407} (\bibinfo {year} {2011})}\BibitemShut {NoStop}%
\bibitem [{\citenamefont {Wang}\ \emph {et~al.}(2017{\natexlab{a}})\citenamefont {Wang}, \citenamefont {Dong}, \citenamefont {Yu},\ and\ \citenamefont {Li}}]{PhysRevB.96.115103}%
  \BibitemOpen
  \bibfield  {author} {\bibinfo {author} {\bibfnamefont {W.}~\bibnamefont {Wang}}, \bibinfo {author} {\bibfnamefont {Z.-Y.}\ \bibnamefont {Dong}}, \bibinfo {author} {\bibfnamefont {S.-L.}\ \bibnamefont {Yu}},\ and\ \bibinfo {author} {\bibfnamefont {J.-X.}\ \bibnamefont {Li}},\ }\href {https://doi.org/10.1103/PhysRevB.96.115103} {\bibfield  {journal} {\bibinfo  {journal} {Phys. Rev. B}\ }\textbf {\bibinfo {volume} {96}},\ \bibinfo {pages} {115103} (\bibinfo {year} {2017}{\natexlab{a}})}\BibitemShut {NoStop}%
\bibitem [{\citenamefont {Janssen}\ \emph {et~al.}(2017)\citenamefont {Janssen}, \citenamefont {Andrade},\ and\ \citenamefont {Vojta}}]{PhysRevB.96.064430}%
  \BibitemOpen
  \bibfield  {author} {\bibinfo {author} {\bibfnamefont {L.}~\bibnamefont {Janssen}}, \bibinfo {author} {\bibfnamefont {E.~C.}\ \bibnamefont {Andrade}},\ and\ \bibinfo {author} {\bibfnamefont {M.}~\bibnamefont {Vojta}},\ }\href {https://doi.org/10.1103/PhysRevB.96.064430} {\bibfield  {journal} {\bibinfo  {journal} {Phys. Rev. B}\ }\textbf {\bibinfo {volume} {96}},\ \bibinfo {pages} {064430} (\bibinfo {year} {2017})}\BibitemShut {NoStop}%
\bibitem [{\citenamefont {Rusna\ifmmode~\check{c}\else \v{c}\fi{}ko}\ \emph {et~al.}(2019)\citenamefont {Rusna\ifmmode~\check{c}\else \v{c}\fi{}ko}, \citenamefont {Gotfryd},\ and\ \citenamefont {Chaloupka}}]{PhysRevB.99.064425}%
  \BibitemOpen
  \bibfield  {author} {\bibinfo {author} {\bibfnamefont {J.}~\bibnamefont {Rusna\ifmmode~\check{c}\else \v{c}\fi{}ko}}, \bibinfo {author} {\bibfnamefont {D.}~\bibnamefont {Gotfryd}},\ and\ \bibinfo {author} {\bibfnamefont {J.~c.~v.}\ \bibnamefont {Chaloupka}},\ }\href {https://doi.org/10.1103/PhysRevB.99.064425} {\bibfield  {journal} {\bibinfo  {journal} {Phys. Rev. B}\ }\textbf {\bibinfo {volume} {99}},\ \bibinfo {pages} {064425} (\bibinfo {year} {2019})}\BibitemShut {NoStop}%
\bibitem [{\citenamefont {Maksimov}\ and\ \citenamefont {Chernyshev}(2020)}]{PhysRevResearch.2.033011}%
  \BibitemOpen
  \bibfield  {author} {\bibinfo {author} {\bibfnamefont {P.~A.}\ \bibnamefont {Maksimov}}\ and\ \bibinfo {author} {\bibfnamefont {A.~L.}\ \bibnamefont {Chernyshev}},\ }\href {https://doi.org/10.1103/PhysRevResearch.2.033011} {\bibfield  {journal} {\bibinfo  {journal} {Phys. Rev. Res.}\ }\textbf {\bibinfo {volume} {2}},\ \bibinfo {pages} {033011} (\bibinfo {year} {2020})}\BibitemShut {NoStop}%
\bibitem [{\citenamefont {Winter}\ \emph {et~al.}(2018)\citenamefont {Winter}, \citenamefont {Riedl}, \citenamefont {Kaib}, \citenamefont {Coldea},\ and\ \citenamefont {Valent\'{\i}}}]{PhysRevLett.120.077203}%
  \BibitemOpen
  \bibfield  {author} {\bibinfo {author} {\bibfnamefont {S.~M.}\ \bibnamefont {Winter}}, \bibinfo {author} {\bibfnamefont {K.}~\bibnamefont {Riedl}}, \bibinfo {author} {\bibfnamefont {D.}~\bibnamefont {Kaib}}, \bibinfo {author} {\bibfnamefont {R.}~\bibnamefont {Coldea}},\ and\ \bibinfo {author} {\bibfnamefont {R.}~\bibnamefont {Valent\'{\i}}},\ }\href {https://doi.org/10.1103/PhysRevLett.120.077203} {\bibfield  {journal} {\bibinfo  {journal} {Phys. Rev. Lett.}\ }\textbf {\bibinfo {volume} {120}},\ \bibinfo {pages} {077203} (\bibinfo {year} {2018})}\BibitemShut {NoStop}%
\bibitem [{\citenamefont {Wang}\ \emph {et~al.}(2018)\citenamefont {Wang}, \citenamefont {Guo}, \citenamefont {Tafti}, \citenamefont {Hegg}, \citenamefont {Sen}, \citenamefont {Sidorov}, \citenamefont {Wang}, \citenamefont {Cai}, \citenamefont {Yi}, \citenamefont {Zhou}, \citenamefont {Wang}, \citenamefont {Zhang}, \citenamefont {Yang}, \citenamefont {Li}, \citenamefont {Li}, \citenamefont {Li}, \citenamefont {Liu}, \citenamefont {Shi}, \citenamefont {Ku}, \citenamefont {Wu}, \citenamefont {Cava},\ and\ \citenamefont {Sun}}]{PhysRevB.97.245149}%
  \BibitemOpen
  \bibfield  {author} {\bibinfo {author} {\bibfnamefont {Z.}~\bibnamefont {Wang}}, \bibinfo {author} {\bibfnamefont {J.}~\bibnamefont {Guo}}, \bibinfo {author} {\bibfnamefont {F.~F.}\ \bibnamefont {Tafti}}, \bibinfo {author} {\bibfnamefont {A.}~\bibnamefont {Hegg}}, \bibinfo {author} {\bibfnamefont {S.}~\bibnamefont {Sen}}, \bibinfo {author} {\bibfnamefont {V.~A.}\ \bibnamefont {Sidorov}}, \bibinfo {author} {\bibfnamefont {L.}~\bibnamefont {Wang}}, \bibinfo {author} {\bibfnamefont {S.}~\bibnamefont {Cai}}, \bibinfo {author} {\bibfnamefont {W.}~\bibnamefont {Yi}}, \bibinfo {author} {\bibfnamefont {Y.}~\bibnamefont {Zhou}}, \bibinfo {author} {\bibfnamefont {H.}~\bibnamefont {Wang}}, \bibinfo {author} {\bibfnamefont {S.}~\bibnamefont {Zhang}}, \bibinfo {author} {\bibfnamefont {K.}~\bibnamefont {Yang}}, \bibinfo {author} {\bibfnamefont {A.}~\bibnamefont {Li}}, \bibinfo {author} {\bibfnamefont {X.}~\bibnamefont {Li}}, \bibinfo {author} {\bibfnamefont {Y.}~\bibnamefont {Li}}, \bibinfo {author} {\bibfnamefont
  {J.}~\bibnamefont {Liu}}, \bibinfo {author} {\bibfnamefont {Y.}~\bibnamefont {Shi}}, \bibinfo {author} {\bibfnamefont {W.}~\bibnamefont {Ku}}, \bibinfo {author} {\bibfnamefont {Q.}~\bibnamefont {Wu}}, \bibinfo {author} {\bibfnamefont {R.~J.}\ \bibnamefont {Cava}},\ and\ \bibinfo {author} {\bibfnamefont {L.}~\bibnamefont {Sun}},\ }\href {https://doi.org/10.1103/PhysRevB.97.245149} {\bibfield  {journal} {\bibinfo  {journal} {Phys. Rev. B}\ }\textbf {\bibinfo {volume} {97}},\ \bibinfo {pages} {245149} (\bibinfo {year} {2018})}\BibitemShut {NoStop}%
\bibitem [{\citenamefont {Liu}\ and\ \citenamefont {Khaliullin}(2018)}]{liu2018pseudospin}%
  \BibitemOpen
  \bibfield  {author} {\bibinfo {author} {\bibfnamefont {H.}~\bibnamefont {Liu}}\ and\ \bibinfo {author} {\bibfnamefont {G.}~\bibnamefont {Khaliullin}},\ }\href {https://doi.org/10.1103/PhysRevB.97.014407} {\bibfield  {journal} {\bibinfo  {journal} {Phys. Rev. B}\ }\textbf {\bibinfo {volume} {97}},\ \bibinfo {pages} {014407} (\bibinfo {year} {2018})}\BibitemShut {NoStop}%
\bibitem [{\citenamefont {Sano}\ \emph {et~al.}(2018)\citenamefont {Sano}, \citenamefont {Kato},\ and\ \citenamefont {Motome}}]{sano2018kitaev}%
  \BibitemOpen
  \bibfield  {author} {\bibinfo {author} {\bibfnamefont {R.}~\bibnamefont {Sano}}, \bibinfo {author} {\bibfnamefont {Y.}~\bibnamefont {Kato}},\ and\ \bibinfo {author} {\bibfnamefont {Y.}~\bibnamefont {Motome}},\ }\href {https://doi.org/10.1103/PhysRevB.97.014408} {\bibfield  {journal} {\bibinfo  {journal} {Phys. Rev. B}\ }\textbf {\bibinfo {volume} {97}},\ \bibinfo {pages} {014408} (\bibinfo {year} {2018})}\BibitemShut {NoStop}%
\bibitem [{\citenamefont {Zhong}\ \emph {et~al.}(2020)\citenamefont {Zhong}, \citenamefont {Gao}, \citenamefont {Ong},\ and\ \citenamefont {Cava}}]{doi:10.1126/sciadv.aay6953}%
  \BibitemOpen
  \bibfield  {author} {\bibinfo {author} {\bibfnamefont {R.}~\bibnamefont {Zhong}}, \bibinfo {author} {\bibfnamefont {T.}~\bibnamefont {Gao}}, \bibinfo {author} {\bibfnamefont {N.~P.}\ \bibnamefont {Ong}},\ and\ \bibinfo {author} {\bibfnamefont {R.~J.}\ \bibnamefont {Cava}},\ }\href {https://doi.org/10.1126/sciadv.aay6953} {\bibfield  {journal} {\bibinfo  {journal} {Science Advances}\ }\textbf {\bibinfo {volume} {6}},\ \bibinfo {pages} {eaay6953} (\bibinfo {year} {2020})},\ \Eprint {https://arxiv.org/abs/https://www.science.org/doi/pdf/10.1126/sciadv.aay6953} {https://www.science.org/doi/pdf/10.1126/sciadv.aay6953} \BibitemShut {NoStop}%
\bibitem [{\citenamefont {Lefran\ifmmode~\mbox{\c{c}}\else \c{c}\fi{}ois}\ \emph {et~al.}(2016)\citenamefont {Lefran\ifmmode~\mbox{\c{c}}\else \c{c}\fi{}ois}, \citenamefont {Songvilay}, \citenamefont {Robert}, \citenamefont {Nataf}, \citenamefont {Jordan}, \citenamefont {Chaix}, \citenamefont {Colin}, \citenamefont {Lejay}, \citenamefont {Hadj-Azzem}, \citenamefont {Ballou},\ and\ \citenamefont {Simonet}}]{PhysRevB.94.214416}%
  \BibitemOpen
  \bibfield  {author} {\bibinfo {author} {\bibfnamefont {E.}~\bibnamefont {Lefran\ifmmode~\mbox{\c{c}}\else \c{c}\fi{}ois}}, \bibinfo {author} {\bibfnamefont {M.}~\bibnamefont {Songvilay}}, \bibinfo {author} {\bibfnamefont {J.}~\bibnamefont {Robert}}, \bibinfo {author} {\bibfnamefont {G.}~\bibnamefont {Nataf}}, \bibinfo {author} {\bibfnamefont {E.}~\bibnamefont {Jordan}}, \bibinfo {author} {\bibfnamefont {L.}~\bibnamefont {Chaix}}, \bibinfo {author} {\bibfnamefont {C.~V.}\ \bibnamefont {Colin}}, \bibinfo {author} {\bibfnamefont {P.}~\bibnamefont {Lejay}}, \bibinfo {author} {\bibfnamefont {A.}~\bibnamefont {Hadj-Azzem}}, \bibinfo {author} {\bibfnamefont {R.}~\bibnamefont {Ballou}},\ and\ \bibinfo {author} {\bibfnamefont {V.}~\bibnamefont {Simonet}},\ }\href {https://doi.org/10.1103/PhysRevB.94.214416} {\bibfield  {journal} {\bibinfo  {journal} {Phys. Rev. B}\ }\textbf {\bibinfo {volume} {94}},\ \bibinfo {pages} {214416} (\bibinfo {year} {2016})}\BibitemShut {NoStop}%
\bibitem [{\citenamefont {Bera}\ \emph {et~al.}(2017)\citenamefont {Bera}, \citenamefont {Yusuf}, \citenamefont {Kumar},\ and\ \citenamefont {Ritter}}]{PhysRevB.95.094424}%
  \BibitemOpen
  \bibfield  {author} {\bibinfo {author} {\bibfnamefont {A.~K.}\ \bibnamefont {Bera}}, \bibinfo {author} {\bibfnamefont {S.~M.}\ \bibnamefont {Yusuf}}, \bibinfo {author} {\bibfnamefont {A.}~\bibnamefont {Kumar}},\ and\ \bibinfo {author} {\bibfnamefont {C.}~\bibnamefont {Ritter}},\ }\href {https://doi.org/10.1103/PhysRevB.95.094424} {\bibfield  {journal} {\bibinfo  {journal} {Phys. Rev. B}\ }\textbf {\bibinfo {volume} {95}},\ \bibinfo {pages} {094424} (\bibinfo {year} {2017})}\BibitemShut {NoStop}%
\bibitem [{\citenamefont {Viciu}\ \emph {et~al.}(2007)\citenamefont {Viciu}, \citenamefont {Huang}, \citenamefont {Morosan}, \citenamefont {Zandbergen}, \citenamefont {Greenbaum}, \citenamefont {McQueen},\ and\ \citenamefont {Cava}}]{VICIU20071060}%
  \BibitemOpen
  \bibfield  {author} {\bibinfo {author} {\bibfnamefont {L.}~\bibnamefont {Viciu}}, \bibinfo {author} {\bibfnamefont {Q.}~\bibnamefont {Huang}}, \bibinfo {author} {\bibfnamefont {E.}~\bibnamefont {Morosan}}, \bibinfo {author} {\bibfnamefont {H.}~\bibnamefont {Zandbergen}}, \bibinfo {author} {\bibfnamefont {N.}~\bibnamefont {Greenbaum}}, \bibinfo {author} {\bibfnamefont {T.}~\bibnamefont {McQueen}},\ and\ \bibinfo {author} {\bibfnamefont {R.}~\bibnamefont {Cava}},\ }\href {https://doi.org/https://doi.org/10.1016/j.jssc.2007.01.002} {\bibfield  {journal} {\bibinfo  {journal} {Journal of Solid State Chemistry}\ }\textbf {\bibinfo {volume} {180}},\ \bibinfo {pages} {1060} (\bibinfo {year} {2007})}\BibitemShut {NoStop}%
\bibitem [{\citenamefont {Liu}\ \emph {et~al.}(2020)\citenamefont {Liu}, \citenamefont {Chaloupka},\ and\ \citenamefont {Khaliullin}}]{PhysRevLett.125.047201}%
  \BibitemOpen
  \bibfield  {author} {\bibinfo {author} {\bibfnamefont {H.}~\bibnamefont {Liu}}, \bibinfo {author} {\bibfnamefont {J.~c.~v.}\ \bibnamefont {Chaloupka}},\ and\ \bibinfo {author} {\bibfnamefont {G.}~\bibnamefont {Khaliullin}},\ }\href {https://doi.org/10.1103/PhysRevLett.125.047201} {\bibfield  {journal} {\bibinfo  {journal} {Phys. Rev. Lett.}\ }\textbf {\bibinfo {volume} {125}},\ \bibinfo {pages} {047201} (\bibinfo {year} {2020})}\BibitemShut {NoStop}%
\bibitem [{\citenamefont {Kim}\ \emph {et~al.}(2021{\natexlab{a}})\citenamefont {Kim}, \citenamefont {Jeong}, \citenamefont {Lin}, \citenamefont {Park}, \citenamefont {Masuda}, \citenamefont {Asai}, \citenamefont {Itoh}, \citenamefont {Kim}, \citenamefont {Zhou}, \citenamefont {Ma},\ and\ \citenamefont {Park}}]{Kim_2022}%
  \BibitemOpen
  \bibfield  {author} {\bibinfo {author} {\bibfnamefont {C.}~\bibnamefont {Kim}}, \bibinfo {author} {\bibfnamefont {J.}~\bibnamefont {Jeong}}, \bibinfo {author} {\bibfnamefont {G.}~\bibnamefont {Lin}}, \bibinfo {author} {\bibfnamefont {P.}~\bibnamefont {Park}}, \bibinfo {author} {\bibfnamefont {T.}~\bibnamefont {Masuda}}, \bibinfo {author} {\bibfnamefont {S.}~\bibnamefont {Asai}}, \bibinfo {author} {\bibfnamefont {S.}~\bibnamefont {Itoh}}, \bibinfo {author} {\bibfnamefont {H.-S.}\ \bibnamefont {Kim}}, \bibinfo {author} {\bibfnamefont {H.}~\bibnamefont {Zhou}}, \bibinfo {author} {\bibfnamefont {J.}~\bibnamefont {Ma}},\ and\ \bibinfo {author} {\bibfnamefont {J.-G.}\ \bibnamefont {Park}},\ }\href {https://doi.org/10.1088/1361-648X/ac2644} {\bibfield  {journal} {\bibinfo  {journal} {Journal of Physics: Condensed Matter}\ }\textbf {\bibinfo {volume} {34}},\ \bibinfo {pages} {045802} (\bibinfo {year} {2021}{\natexlab{a}})}\BibitemShut {NoStop}%
\bibitem [{\citenamefont {Vavilova}\ \emph {et~al.}(2023)\citenamefont {Vavilova}, \citenamefont {Vasilchikova}, \citenamefont {Vasiliev}, \citenamefont {Mikhailova}, \citenamefont {Nalbandyan}, \citenamefont {Zvereva},\ and\ \citenamefont {Streltsov}}]{PhysRevB.107.054411}%
  \BibitemOpen
  \bibfield  {author} {\bibinfo {author} {\bibfnamefont {E.}~\bibnamefont {Vavilova}}, \bibinfo {author} {\bibfnamefont {T.}~\bibnamefont {Vasilchikova}}, \bibinfo {author} {\bibfnamefont {A.}~\bibnamefont {Vasiliev}}, \bibinfo {author} {\bibfnamefont {D.}~\bibnamefont {Mikhailova}}, \bibinfo {author} {\bibfnamefont {V.}~\bibnamefont {Nalbandyan}}, \bibinfo {author} {\bibfnamefont {E.}~\bibnamefont {Zvereva}},\ and\ \bibinfo {author} {\bibfnamefont {S.~V.}\ \bibnamefont {Streltsov}},\ }\href {https://doi.org/10.1103/PhysRevB.107.054411} {\bibfield  {journal} {\bibinfo  {journal} {Phys. Rev. B}\ }\textbf {\bibinfo {volume} {107}},\ \bibinfo {pages} {054411} (\bibinfo {year} {2023})}\BibitemShut {NoStop}%
\bibitem [{\citenamefont {Wong}\ \emph {et~al.}(2016)\citenamefont {Wong}, \citenamefont {Avdeev},\ and\ \citenamefont {Ling}}]{WONG201618}%
  \BibitemOpen
  \bibfield  {author} {\bibinfo {author} {\bibfnamefont {C.}~\bibnamefont {Wong}}, \bibinfo {author} {\bibfnamefont {M.}~\bibnamefont {Avdeev}},\ and\ \bibinfo {author} {\bibfnamefont {C.~D.}\ \bibnamefont {Ling}},\ }\href {https://doi.org/https://doi.org/10.1016/j.jssc.2016.07.032} {\bibfield  {journal} {\bibinfo  {journal} {Journal of Solid State Chemistry}\ }\textbf {\bibinfo {volume} {243}},\ \bibinfo {pages} {18} (\bibinfo {year} {2016})}\BibitemShut {NoStop}%
\bibitem [{\citenamefont {Yan}\ \emph {et~al.}(2019)\citenamefont {Yan}, \citenamefont {Okamoto}, \citenamefont {Wu}, \citenamefont {Zheng}, \citenamefont {Zhou}, \citenamefont {Cao},\ and\ \citenamefont {McGuire}}]{PhysRevMaterials.3.074405}%
  \BibitemOpen
  \bibfield  {author} {\bibinfo {author} {\bibfnamefont {J.-Q.}\ \bibnamefont {Yan}}, \bibinfo {author} {\bibfnamefont {S.}~\bibnamefont {Okamoto}}, \bibinfo {author} {\bibfnamefont {Y.}~\bibnamefont {Wu}}, \bibinfo {author} {\bibfnamefont {Q.}~\bibnamefont {Zheng}}, \bibinfo {author} {\bibfnamefont {H.~D.}\ \bibnamefont {Zhou}}, \bibinfo {author} {\bibfnamefont {H.~B.}\ \bibnamefont {Cao}},\ and\ \bibinfo {author} {\bibfnamefont {M.~A.}\ \bibnamefont {McGuire}},\ }\href {https://doi.org/10.1103/PhysRevMaterials.3.074405} {\bibfield  {journal} {\bibinfo  {journal} {Phys. Rev. Mater.}\ }\textbf {\bibinfo {volume} {3}},\ \bibinfo {pages} {074405} (\bibinfo {year} {2019})}\BibitemShut {NoStop}%
\bibitem [{\citenamefont {Stratan}\ \emph {et~al.}(2019)\citenamefont {Stratan}, \citenamefont {Shukaev}, \citenamefont {Vasilchikova}, \citenamefont {Vasiliev}, \citenamefont {Korshunov}, \citenamefont {Kurbakov}, \citenamefont {Nalbandyan},\ and\ \citenamefont {Zvereva}}]{C9NJ03627J}%
  \BibitemOpen
  \bibfield  {author} {\bibinfo {author} {\bibfnamefont {M.~I.}\ \bibnamefont {Stratan}}, \bibinfo {author} {\bibfnamefont {I.~L.}\ \bibnamefont {Shukaev}}, \bibinfo {author} {\bibfnamefont {T.~M.}\ \bibnamefont {Vasilchikova}}, \bibinfo {author} {\bibfnamefont {A.~N.}\ \bibnamefont {Vasiliev}}, \bibinfo {author} {\bibfnamefont {A.~N.}\ \bibnamefont {Korshunov}}, \bibinfo {author} {\bibfnamefont {A.~I.}\ \bibnamefont {Kurbakov}}, \bibinfo {author} {\bibfnamefont {V.~B.}\ \bibnamefont {Nalbandyan}},\ and\ \bibinfo {author} {\bibfnamefont {E.~A.}\ \bibnamefont {Zvereva}},\ }\href {https://doi.org/10.1039/C9NJ03627J} {\bibfield  {journal} {\bibinfo  {journal} {New J. Chem.}\ }\textbf {\bibinfo {volume} {43}},\ \bibinfo {pages} {13545} (\bibinfo {year} {2019})}\BibitemShut {NoStop}%
\bibitem [{\citenamefont {Songvilay}\ \emph {et~al.}(2020)\citenamefont {Songvilay}, \citenamefont {Robert}, \citenamefont {Petit}, \citenamefont {Rodriguez-Rivera}, \citenamefont {Ratcliff}, \citenamefont {Damay}, \citenamefont {Bal\'edent}, \citenamefont {Jim\'enez-Ruiz}, \citenamefont {Lejay}, \citenamefont {Pachoud}, \citenamefont {Hadj-Azzem}, \citenamefont {Simonet},\ and\ \citenamefont {Stock}}]{NSCO_neutron_Kitaev_parameter}%
  \BibitemOpen
  \bibfield  {author} {\bibinfo {author} {\bibfnamefont {M.}~\bibnamefont {Songvilay}}, \bibinfo {author} {\bibfnamefont {J.}~\bibnamefont {Robert}}, \bibinfo {author} {\bibfnamefont {S.}~\bibnamefont {Petit}}, \bibinfo {author} {\bibfnamefont {J.~A.}\ \bibnamefont {Rodriguez-Rivera}}, \bibinfo {author} {\bibfnamefont {W.~D.}\ \bibnamefont {Ratcliff}}, \bibinfo {author} {\bibfnamefont {F.}~\bibnamefont {Damay}}, \bibinfo {author} {\bibfnamefont {V.}~\bibnamefont {Bal\'edent}}, \bibinfo {author} {\bibfnamefont {M.}~\bibnamefont {Jim\'enez-Ruiz}}, \bibinfo {author} {\bibfnamefont {P.}~\bibnamefont {Lejay}}, \bibinfo {author} {\bibfnamefont {E.}~\bibnamefont {Pachoud}}, \bibinfo {author} {\bibfnamefont {A.}~\bibnamefont {Hadj-Azzem}}, \bibinfo {author} {\bibfnamefont {V.}~\bibnamefont {Simonet}},\ and\ \bibinfo {author} {\bibfnamefont {C.}~\bibnamefont {Stock}},\ }\href {https://doi.org/10.1103/PhysRevB.102.224429} {\bibfield  {journal} {\bibinfo  {journal} {Phys. Rev. B}\ }\textbf {\bibinfo {volume} {102}},\
  \bibinfo {pages} {224429} (\bibinfo {year} {2020})}\BibitemShut {NoStop}%
\bibitem [{\citenamefont {Kim}\ \emph {et~al.}(2021{\natexlab{b}})\citenamefont {Kim}, \citenamefont {Kim},\ and\ \citenamefont {Park}}]{Kim_3d}%
  \BibitemOpen
  \bibfield  {author} {\bibinfo {author} {\bibfnamefont {C.}~\bibnamefont {Kim}}, \bibinfo {author} {\bibfnamefont {H.-S.}\ \bibnamefont {Kim}},\ and\ \bibinfo {author} {\bibfnamefont {J.-G.}\ \bibnamefont {Park}},\ }\href {https://doi.org/10.1088/1361-648X/ac2d5d} {\bibfield  {journal} {\bibinfo  {journal} {Journal of Physics: Condensed Matter}\ }\textbf {\bibinfo {volume} {34}},\ \bibinfo {pages} {023001} (\bibinfo {year} {2021}{\natexlab{b}})}\BibitemShut {NoStop}%
\bibitem [{\citenamefont {Gu}\ \emph {et~al.}(2024)\citenamefont {Gu}, \citenamefont {Li}, \citenamefont {Chen}, \citenamefont {Iida}, \citenamefont {Nakao}, \citenamefont {Munakata}, \citenamefont {Garlea}, \citenamefont {Li}, \citenamefont {Deng}, \citenamefont {Zaliznyak}, \citenamefont {Tranquada},\ and\ \citenamefont {Li}}]{Gu_NCSO_double_Q}%
  \BibitemOpen
  \bibfield  {author} {\bibinfo {author} {\bibfnamefont {Y.}~\bibnamefont {Gu}}, \bibinfo {author} {\bibfnamefont {X.}~\bibnamefont {Li}}, \bibinfo {author} {\bibfnamefont {Y.}~\bibnamefont {Chen}}, \bibinfo {author} {\bibfnamefont {K.}~\bibnamefont {Iida}}, \bibinfo {author} {\bibfnamefont {A.}~\bibnamefont {Nakao}}, \bibinfo {author} {\bibfnamefont {K.}~\bibnamefont {Munakata}}, \bibinfo {author} {\bibfnamefont {V.~O.}\ \bibnamefont {Garlea}}, \bibinfo {author} {\bibfnamefont {Y.}~\bibnamefont {Li}}, \bibinfo {author} {\bibfnamefont {G.}~\bibnamefont {Deng}}, \bibinfo {author} {\bibfnamefont {I.~A.}\ \bibnamefont {Zaliznyak}}, \bibinfo {author} {\bibfnamefont {J.~M.}\ \bibnamefont {Tranquada}},\ and\ \bibinfo {author} {\bibfnamefont {Y.}~\bibnamefont {Li}},\ }\href {https://doi.org/10.1103/PhysRevB.109.L060410} {\bibfield  {journal} {\bibinfo  {journal} {Phys. Rev. B}\ }\textbf {\bibinfo {volume} {109}},\ \bibinfo {pages} {L060410} (\bibinfo {year} {2024})}\BibitemShut {NoStop}%
\bibitem [{\citenamefont {Merino}\ and\ \citenamefont {Ralko}(2018)}]{QSL_XXZ_model}%
  \BibitemOpen
  \bibfield  {author} {\bibinfo {author} {\bibfnamefont {J.}~\bibnamefont {Merino}}\ and\ \bibinfo {author} {\bibfnamefont {A.}~\bibnamefont {Ralko}},\ }\href {https://doi.org/10.1103/PhysRevB.97.205112} {\bibfield  {journal} {\bibinfo  {journal} {Phys. Rev. B}\ }\textbf {\bibinfo {volume} {97}},\ \bibinfo {pages} {205112} (\bibinfo {year} {2018})}\BibitemShut {NoStop}%
\bibitem [{\citenamefont {Bose}\ \emph {et~al.}(2023)\citenamefont {Bose}, \citenamefont {Routh}, \citenamefont {Voleti}, \citenamefont {Saha}, \citenamefont {Kumar}, \citenamefont {Saha-Dasgupta},\ and\ \citenamefont {Paramekanti}}]{XXZ_model1}%
  \BibitemOpen
  \bibfield  {author} {\bibinfo {author} {\bibfnamefont {A.}~\bibnamefont {Bose}}, \bibinfo {author} {\bibfnamefont {M.}~\bibnamefont {Routh}}, \bibinfo {author} {\bibfnamefont {S.}~\bibnamefont {Voleti}}, \bibinfo {author} {\bibfnamefont {S.~K.}\ \bibnamefont {Saha}}, \bibinfo {author} {\bibfnamefont {M.}~\bibnamefont {Kumar}}, \bibinfo {author} {\bibfnamefont {T.}~\bibnamefont {Saha-Dasgupta}},\ and\ \bibinfo {author} {\bibfnamefont {A.}~\bibnamefont {Paramekanti}},\ }\href {https://doi.org/10.1103/PhysRevB.108.174422} {\bibfield  {journal} {\bibinfo  {journal} {Phys. Rev. B}\ }\textbf {\bibinfo {volume} {108}},\ \bibinfo {pages} {174422} (\bibinfo {year} {2023})}\BibitemShut {NoStop}%
\bibitem [{\citenamefont {Li}\ \emph {et~al.}(2022)\citenamefont {Li}, \citenamefont {Gu}, \citenamefont {Chen}, \citenamefont {Garlea}, \citenamefont {Iida}, \citenamefont {Kamazawa}, \citenamefont {Li}, \citenamefont {Deng}, \citenamefont {Xiao}, \citenamefont {Zheng}, \citenamefont {Ye}, \citenamefont {Peng}, \citenamefont {Zaliznyak}, \citenamefont {Tranquada},\ and\ \citenamefont {Li}}]{li2022giant}%
  \BibitemOpen
  \bibfield  {author} {\bibinfo {author} {\bibfnamefont {X.}~\bibnamefont {Li}}, \bibinfo {author} {\bibfnamefont {Y.}~\bibnamefont {Gu}}, \bibinfo {author} {\bibfnamefont {Y.}~\bibnamefont {Chen}}, \bibinfo {author} {\bibfnamefont {V.~O.}\ \bibnamefont {Garlea}}, \bibinfo {author} {\bibfnamefont {K.}~\bibnamefont {Iida}}, \bibinfo {author} {\bibfnamefont {K.}~\bibnamefont {Kamazawa}}, \bibinfo {author} {\bibfnamefont {Y.}~\bibnamefont {Li}}, \bibinfo {author} {\bibfnamefont {G.}~\bibnamefont {Deng}}, \bibinfo {author} {\bibfnamefont {Q.}~\bibnamefont {Xiao}}, \bibinfo {author} {\bibfnamefont {X.}~\bibnamefont {Zheng}}, \bibinfo {author} {\bibfnamefont {Z.}~\bibnamefont {Ye}}, \bibinfo {author} {\bibfnamefont {Y.}~\bibnamefont {Peng}}, \bibinfo {author} {\bibfnamefont {I.~A.}\ \bibnamefont {Zaliznyak}}, \bibinfo {author} {\bibfnamefont {J.~M.}\ \bibnamefont {Tranquada}},\ and\ \bibinfo {author} {\bibfnamefont {Y.}~\bibnamefont {Li}},\ }\href {https://doi.org/10.1103/PhysRevX.12.041024} {\bibfield  {journal}
  {\bibinfo  {journal} {Phys. Rev. X}\ }\textbf {\bibinfo {volume} {12}},\ \bibinfo {pages} {041024} (\bibinfo {year} {2022})}\BibitemShut {NoStop}%
\bibitem [{\citenamefont {Hu}\ \emph {et~al.}(2024)\citenamefont {Hu}, \citenamefont {Chen}, \citenamefont {Cui}, \citenamefont {Li}, \citenamefont {Li}, \citenamefont {Xu}, \citenamefont {Chen}, \citenamefont {Li}, \citenamefont {Gu}, \citenamefont {Yu}, \citenamefont {Zhou}, \citenamefont {Li},\ and\ \citenamefont {Yu}}]{PhysRevB.109.054411}%
  \BibitemOpen
  \bibfield  {author} {\bibinfo {author} {\bibfnamefont {Z.}~\bibnamefont {Hu}}, \bibinfo {author} {\bibfnamefont {Y.}~\bibnamefont {Chen}}, \bibinfo {author} {\bibfnamefont {Y.}~\bibnamefont {Cui}}, \bibinfo {author} {\bibfnamefont {S.}~\bibnamefont {Li}}, \bibinfo {author} {\bibfnamefont {C.}~\bibnamefont {Li}}, \bibinfo {author} {\bibfnamefont {X.}~\bibnamefont {Xu}}, \bibinfo {author} {\bibfnamefont {Y.}~\bibnamefont {Chen}}, \bibinfo {author} {\bibfnamefont {X.}~\bibnamefont {Li}}, \bibinfo {author} {\bibfnamefont {Y.}~\bibnamefont {Gu}}, \bibinfo {author} {\bibfnamefont {R.}~\bibnamefont {Yu}}, \bibinfo {author} {\bibfnamefont {R.}~\bibnamefont {Zhou}}, \bibinfo {author} {\bibfnamefont {Y.}~\bibnamefont {Li}},\ and\ \bibinfo {author} {\bibfnamefont {W.}~\bibnamefont {Yu}},\ }\href {https://doi.org/10.1103/PhysRevB.109.054411} {\bibfield  {journal} {\bibinfo  {journal} {Phys. Rev. B}\ }\textbf {\bibinfo {volume} {109}},\ \bibinfo {pages} {054411} (\bibinfo {year} {2024})}\BibitemShut {NoStop}%
\bibitem [{\citenamefont {Sheng}\ \emph {et~al.}(2025)\citenamefont {Sheng}, \citenamefont {Wang}, \citenamefont {Jiang}, \citenamefont {Ge}, \citenamefont {Zhao}, \citenamefont {Li}, \citenamefont {Kofu}, \citenamefont {Yu}, \citenamefont {Zhu}, \citenamefont {Mei}, \citenamefont {Wang},\ and\ \citenamefont {Wu}}]{NaBaCoPO_neutron}%
  \BibitemOpen
  \bibfield  {author} {\bibinfo {author} {\bibfnamefont {J.}~\bibnamefont {Sheng}}, \bibinfo {author} {\bibfnamefont {L.}~\bibnamefont {Wang}}, \bibinfo {author} {\bibfnamefont {W.}~\bibnamefont {Jiang}}, \bibinfo {author} {\bibfnamefont {H.}~\bibnamefont {Ge}}, \bibinfo {author} {\bibfnamefont {N.}~\bibnamefont {Zhao}}, \bibinfo {author} {\bibfnamefont {T.}~\bibnamefont {Li}}, \bibinfo {author} {\bibfnamefont {M.}~\bibnamefont {Kofu}}, \bibinfo {author} {\bibfnamefont {D.}~\bibnamefont {Yu}}, \bibinfo {author} {\bibfnamefont {W.}~\bibnamefont {Zhu}}, \bibinfo {author} {\bibfnamefont {J.-W.}\ \bibnamefont {Mei}}, \bibinfo {author} {\bibfnamefont {Z.}~\bibnamefont {Wang}},\ and\ \bibinfo {author} {\bibfnamefont {L.}~\bibnamefont {Wu}},\ }\bibfield  {journal} {\bibinfo  {journal} {The Innovation}\ }\textbf {\bibinfo {volume} {6}},\ \href {https://doi.org/10.1016/j.xinn.2024.100769} {10.1016/j.xinn.2024.100769} (\bibinfo {year} {2025})\BibitemShut {NoStop}%
\bibitem [{\citenamefont {Ma}\ \emph {et~al.}(2018)\citenamefont {Ma}, \citenamefont {Wang}, \citenamefont {Dong}, \citenamefont {Zhang}, \citenamefont {Li}, \citenamefont {Zheng}, \citenamefont {Yu}, \citenamefont {Wang}, \citenamefont {Che}, \citenamefont {Ran}, \citenamefont {Bao}, \citenamefont {Cai}, \citenamefont {\ifmmode~\check{C}\else \v{C}\fi{}erm\'ak}, \citenamefont {Schneidewind}, \citenamefont {Yano}, \citenamefont {Gardner}, \citenamefont {Lu}, \citenamefont {Yu}, \citenamefont {Liu}, \citenamefont {Li}, \citenamefont {Li},\ and\ \citenamefont {Wen}}]{Disorder_QSL}%
  \BibitemOpen
  \bibfield  {author} {\bibinfo {author} {\bibfnamefont {Z.}~\bibnamefont {Ma}}, \bibinfo {author} {\bibfnamefont {J.}~\bibnamefont {Wang}}, \bibinfo {author} {\bibfnamefont {Z.-Y.}\ \bibnamefont {Dong}}, \bibinfo {author} {\bibfnamefont {J.}~\bibnamefont {Zhang}}, \bibinfo {author} {\bibfnamefont {S.}~\bibnamefont {Li}}, \bibinfo {author} {\bibfnamefont {S.-H.}\ \bibnamefont {Zheng}}, \bibinfo {author} {\bibfnamefont {Y.}~\bibnamefont {Yu}}, \bibinfo {author} {\bibfnamefont {W.}~\bibnamefont {Wang}}, \bibinfo {author} {\bibfnamefont {L.}~\bibnamefont {Che}}, \bibinfo {author} {\bibfnamefont {K.}~\bibnamefont {Ran}}, \bibinfo {author} {\bibfnamefont {S.}~\bibnamefont {Bao}}, \bibinfo {author} {\bibfnamefont {Z.}~\bibnamefont {Cai}}, \bibinfo {author} {\bibfnamefont {P.}~\bibnamefont {\ifmmode~\check{C}\else \v{C}\fi{}erm\'ak}}, \bibinfo {author} {\bibfnamefont {A.}~\bibnamefont {Schneidewind}}, \bibinfo {author} {\bibfnamefont {S.}~\bibnamefont {Yano}}, \bibinfo {author} {\bibfnamefont {J.~S.}\ \bibnamefont
  {Gardner}}, \bibinfo {author} {\bibfnamefont {X.}~\bibnamefont {Lu}}, \bibinfo {author} {\bibfnamefont {S.-L.}\ \bibnamefont {Yu}}, \bibinfo {author} {\bibfnamefont {J.-M.}\ \bibnamefont {Liu}}, \bibinfo {author} {\bibfnamefont {S.}~\bibnamefont {Li}}, \bibinfo {author} {\bibfnamefont {J.-X.}\ \bibnamefont {Li}},\ and\ \bibinfo {author} {\bibfnamefont {J.}~\bibnamefont {Wen}},\ }\href {https://doi.org/10.1103/PhysRevLett.120.087201} {\bibfield  {journal} {\bibinfo  {journal} {Phys. Rev. Lett.}\ }\textbf {\bibinfo {volume} {120}},\ \bibinfo {pages} {087201} (\bibinfo {year} {2018})}\BibitemShut {NoStop}%
\bibitem [{\citenamefont {Han}\ \emph {et~al.}(2012)\citenamefont {Han}, \citenamefont {Helton}, \citenamefont {Chu}, \citenamefont {Nocera}, \citenamefont {Rodriguez-Rivera}, \citenamefont {Broholm},\ and\ \citenamefont {Lee}}]{continuum_neutron_Cu}%
  \BibitemOpen
  \bibfield  {author} {\bibinfo {author} {\bibfnamefont {T.-H.}\ \bibnamefont {Han}}, \bibinfo {author} {\bibfnamefont {J.~S.}\ \bibnamefont {Helton}}, \bibinfo {author} {\bibfnamefont {S.}~\bibnamefont {Chu}}, \bibinfo {author} {\bibfnamefont {D.~G.}\ \bibnamefont {Nocera}}, \bibinfo {author} {\bibfnamefont {J.~A.}\ \bibnamefont {Rodriguez-Rivera}}, \bibinfo {author} {\bibfnamefont {C.}~\bibnamefont {Broholm}},\ and\ \bibinfo {author} {\bibfnamefont {Y.~S.}\ \bibnamefont {Lee}},\ }\href {https://doi.org/10.1038/nature11659} {\bibfield  {journal} {\bibinfo  {journal} {Nature}\ }\textbf {\bibinfo {volume} {492}},\ \bibinfo {pages} {406} (\bibinfo {year} {2012})}\BibitemShut {NoStop}%
\bibitem [{sup()}]{supplemental_material}%
  \BibitemOpen
  \href@noop {} {\bibinfo {title} {See {Supplemental Material} for sample characterization, methods.}}\BibitemShut {Stop}%
\bibitem [{\citenamefont {Zhang}\ \emph {et~al.}(2023)\citenamefont {Zhang}, \citenamefont {Xu}, \citenamefont {Halloran}, \citenamefont {Zhong}, \citenamefont {Broholm}, \citenamefont {Cava}, \citenamefont {Drichko},\ and\ \citenamefont {Armitage}}]{zhang2023magnetic}%
  \BibitemOpen
  \bibfield  {author} {\bibinfo {author} {\bibfnamefont {X.}~\bibnamefont {Zhang}}, \bibinfo {author} {\bibfnamefont {Y.}~\bibnamefont {Xu}}, \bibinfo {author} {\bibfnamefont {T.}~\bibnamefont {Halloran}}, \bibinfo {author} {\bibfnamefont {R.}~\bibnamefont {Zhong}}, \bibinfo {author} {\bibfnamefont {C.}~\bibnamefont {Broholm}}, \bibinfo {author} {\bibfnamefont {R.}~\bibnamefont {Cava}}, \bibinfo {author} {\bibfnamefont {N.}~\bibnamefont {Drichko}},\ and\ \bibinfo {author} {\bibfnamefont {N.}~\bibnamefont {Armitage}},\ }\href {https://doi.org/10.1038/s41563-022-01403-1} {\bibfield  {journal} {\bibinfo  {journal} {Nature Materials}\ }\textbf {\bibinfo {volume} {22}},\ \bibinfo {pages} {58} (\bibinfo {year} {2023})}\BibitemShut {NoStop}%
\bibitem [{\citenamefont {Knolle}\ \emph {et~al.}(2014{\natexlab{b}})\citenamefont {Knolle}, \citenamefont {Kovrizhin}, \citenamefont {Chalker},\ and\ \citenamefont {Moessner}}]{Knolle_theory_kitaev}%
  \BibitemOpen
  \bibfield  {author} {\bibinfo {author} {\bibfnamefont {J.}~\bibnamefont {Knolle}}, \bibinfo {author} {\bibfnamefont {D.~L.}\ \bibnamefont {Kovrizhin}}, \bibinfo {author} {\bibfnamefont {J.~T.}\ \bibnamefont {Chalker}},\ and\ \bibinfo {author} {\bibfnamefont {R.}~\bibnamefont {Moessner}},\ }\href {https://doi.org/10.1103/PhysRevLett.112.207203} {\bibfield  {journal} {\bibinfo  {journal} {Phys. Rev. Lett.}\ }\textbf {\bibinfo {volume} {112}},\ \bibinfo {pages} {207203} (\bibinfo {year} {2014}{\natexlab{b}})}\BibitemShut {NoStop}%
\bibitem [{\citenamefont {Little}\ \emph {et~al.}(2017)\citenamefont {Little}, \citenamefont {Wu}, \citenamefont {Lampen-Kelley}, \citenamefont {Banerjee}, \citenamefont {Patankar}, \citenamefont {Rees}, \citenamefont {Bridges}, \citenamefont {Yan}, \citenamefont {Mandrus}, \citenamefont {Nagler},\ and\ \citenamefont {Orenstein}}]{PhysRevLett.119.227201}%
  \BibitemOpen
  \bibfield  {author} {\bibinfo {author} {\bibfnamefont {A.}~\bibnamefont {Little}}, \bibinfo {author} {\bibfnamefont {L.}~\bibnamefont {Wu}}, \bibinfo {author} {\bibfnamefont {P.}~\bibnamefont {Lampen-Kelley}}, \bibinfo {author} {\bibfnamefont {A.}~\bibnamefont {Banerjee}}, \bibinfo {author} {\bibfnamefont {S.}~\bibnamefont {Patankar}}, \bibinfo {author} {\bibfnamefont {D.}~\bibnamefont {Rees}}, \bibinfo {author} {\bibfnamefont {C.~A.}\ \bibnamefont {Bridges}}, \bibinfo {author} {\bibfnamefont {J.-Q.}\ \bibnamefont {Yan}}, \bibinfo {author} {\bibfnamefont {D.}~\bibnamefont {Mandrus}}, \bibinfo {author} {\bibfnamefont {S.~E.}\ \bibnamefont {Nagler}},\ and\ \bibinfo {author} {\bibfnamefont {J.}~\bibnamefont {Orenstein}},\ }\href {https://doi.org/10.1103/PhysRevLett.119.227201} {\bibfield  {journal} {\bibinfo  {journal} {Phys. Rev. Lett.}\ }\textbf {\bibinfo {volume} {119}},\ \bibinfo {pages} {227201} (\bibinfo {year} {2017})}\BibitemShut {NoStop}%
\bibitem [{\citenamefont {Sears}\ \emph {et~al.}(2017)\citenamefont {Sears}, \citenamefont {Zhao}, \citenamefont {Xu}, \citenamefont {Lynn},\ and\ \citenamefont {Kim}}]{RuCl3_domains}%
  \BibitemOpen
  \bibfield  {author} {\bibinfo {author} {\bibfnamefont {J.~A.}\ \bibnamefont {Sears}}, \bibinfo {author} {\bibfnamefont {Y.}~\bibnamefont {Zhao}}, \bibinfo {author} {\bibfnamefont {Z.}~\bibnamefont {Xu}}, \bibinfo {author} {\bibfnamefont {J.~W.}\ \bibnamefont {Lynn}},\ and\ \bibinfo {author} {\bibfnamefont {Y.-J.}\ \bibnamefont {Kim}},\ }\href {https://doi.org/10.1103/PhysRevB.95.180411} {\bibfield  {journal} {\bibinfo  {journal} {Phys. Rev. B}\ }\textbf {\bibinfo {volume} {95}},\ \bibinfo {pages} {180411} (\bibinfo {year} {2017})}\BibitemShut {NoStop}%
\bibitem [{\citenamefont {Shi}\ \emph {et~al.}(2021)\citenamefont {Shi}, \citenamefont {Wang}, \citenamefont {Zhong}, \citenamefont {Wang}, \citenamefont {Hu}, \citenamefont {Zhang}, \citenamefont {Liu}, \citenamefont {Dong}, \citenamefont {Wang},\ and\ \citenamefont {Wang}}]{Shi2021}%
  \BibitemOpen
  \bibfield  {author} {\bibinfo {author} {\bibfnamefont {L.~Y.}\ \bibnamefont {Shi}}, \bibinfo {author} {\bibfnamefont {X.~M.}\ \bibnamefont {Wang}}, \bibinfo {author} {\bibfnamefont {R.~D.}\ \bibnamefont {Zhong}}, \bibinfo {author} {\bibfnamefont {Z.~X.}\ \bibnamefont {Wang}}, \bibinfo {author} {\bibfnamefont {T.~C.}\ \bibnamefont {Hu}}, \bibinfo {author} {\bibfnamefont {S.~J.}\ \bibnamefont {Zhang}}, \bibinfo {author} {\bibfnamefont {Q.~M.}\ \bibnamefont {Liu}}, \bibinfo {author} {\bibfnamefont {T.}~\bibnamefont {Dong}}, \bibinfo {author} {\bibfnamefont {F.}~\bibnamefont {Wang}},\ and\ \bibinfo {author} {\bibfnamefont {N.~L.}\ \bibnamefont {Wang}},\ }\href {https://doi.org/10.1103/PhysRevB.104.144408} {\bibfield  {journal} {\bibinfo  {journal} {Phys. Rev. B}\ }\textbf {\bibinfo {volume} {104}},\ \bibinfo {pages} {144408} (\bibinfo {year} {2021})}\BibitemShut {NoStop}%
\bibitem [{\citenamefont {Shi}\ \emph {et~al.}(2018)\citenamefont {Shi}, \citenamefont {Liu}, \citenamefont {Lin}, \citenamefont {Zhang}, \citenamefont {Zhang}, \citenamefont {Wang}, \citenamefont {Shi}, \citenamefont {Dong},\ and\ \citenamefont {Wang}}]{shi2018field}%
  \BibitemOpen
  \bibfield  {author} {\bibinfo {author} {\bibfnamefont {L.~Y.}\ \bibnamefont {Shi}}, \bibinfo {author} {\bibfnamefont {Y.~Q.}\ \bibnamefont {Liu}}, \bibinfo {author} {\bibfnamefont {T.}~\bibnamefont {Lin}}, \bibinfo {author} {\bibfnamefont {M.~Y.}\ \bibnamefont {Zhang}}, \bibinfo {author} {\bibfnamefont {S.~J.}\ \bibnamefont {Zhang}}, \bibinfo {author} {\bibfnamefont {L.}~\bibnamefont {Wang}}, \bibinfo {author} {\bibfnamefont {Y.~G.}\ \bibnamefont {Shi}}, \bibinfo {author} {\bibfnamefont {T.}~\bibnamefont {Dong}},\ and\ \bibinfo {author} {\bibfnamefont {N.~L.}\ \bibnamefont {Wang}},\ }\href {https://doi.org/10.1103/PhysRevB.98.094414} {\bibfield  {journal} {\bibinfo  {journal} {Phys. Rev. B}\ }\textbf {\bibinfo {volume} {98}},\ \bibinfo {pages} {094414} (\bibinfo {year} {2018})}\BibitemShut {NoStop}%
\bibitem [{\citenamefont {Wu}\ \emph {et~al.}(2018)\citenamefont {Wu}, \citenamefont {Little}, \citenamefont {Aldape}, \citenamefont {Rees}, \citenamefont {Thewalt}, \citenamefont {Lampen-Kelley}, \citenamefont {Banerjee}, \citenamefont {Bridges}, \citenamefont {Yan}, \citenamefont {Boone}, \citenamefont {Patankar}, \citenamefont {Goldhaber-Gordon}, \citenamefont {Mandrus}, \citenamefont {Nagler}, \citenamefont {Altman},\ and\ \citenamefont {Orenstein}}]{PhysRevB.98.094425}%
  \BibitemOpen
  \bibfield  {author} {\bibinfo {author} {\bibfnamefont {L.}~\bibnamefont {Wu}}, \bibinfo {author} {\bibfnamefont {A.}~\bibnamefont {Little}}, \bibinfo {author} {\bibfnamefont {E.~E.}\ \bibnamefont {Aldape}}, \bibinfo {author} {\bibfnamefont {D.}~\bibnamefont {Rees}}, \bibinfo {author} {\bibfnamefont {E.}~\bibnamefont {Thewalt}}, \bibinfo {author} {\bibfnamefont {P.}~\bibnamefont {Lampen-Kelley}}, \bibinfo {author} {\bibfnamefont {A.}~\bibnamefont {Banerjee}}, \bibinfo {author} {\bibfnamefont {C.~A.}\ \bibnamefont {Bridges}}, \bibinfo {author} {\bibfnamefont {J.-Q.}\ \bibnamefont {Yan}}, \bibinfo {author} {\bibfnamefont {D.}~\bibnamefont {Boone}}, \bibinfo {author} {\bibfnamefont {S.}~\bibnamefont {Patankar}}, \bibinfo {author} {\bibfnamefont {D.}~\bibnamefont {Goldhaber-Gordon}}, \bibinfo {author} {\bibfnamefont {D.}~\bibnamefont {Mandrus}}, \bibinfo {author} {\bibfnamefont {S.~E.}\ \bibnamefont {Nagler}}, \bibinfo {author} {\bibfnamefont {E.}~\bibnamefont {Altman}},\ and\ \bibinfo {author} {\bibfnamefont
  {J.}~\bibnamefont {Orenstein}},\ }\href {https://doi.org/10.1103/PhysRevB.98.094425} {\bibfield  {journal} {\bibinfo  {journal} {Phys. Rev. B}\ }\textbf {\bibinfo {volume} {98}},\ \bibinfo {pages} {094425} (\bibinfo {year} {2018})}\BibitemShut {NoStop}%
\bibitem [{\citenamefont {Sandilands}\ \emph {et~al.}(2015)\citenamefont {Sandilands}, \citenamefont {Tian}, \citenamefont {Plumb}, \citenamefont {Kim},\ and\ \citenamefont {Burch}}]{PhysRevLett.114.147201}%
  \BibitemOpen
  \bibfield  {author} {\bibinfo {author} {\bibfnamefont {L.~J.}\ \bibnamefont {Sandilands}}, \bibinfo {author} {\bibfnamefont {Y.}~\bibnamefont {Tian}}, \bibinfo {author} {\bibfnamefont {K.~W.}\ \bibnamefont {Plumb}}, \bibinfo {author} {\bibfnamefont {Y.-J.}\ \bibnamefont {Kim}},\ and\ \bibinfo {author} {\bibfnamefont {K.~S.}\ \bibnamefont {Burch}},\ }\href {https://doi.org/10.1103/PhysRevLett.114.147201} {\bibfield  {journal} {\bibinfo  {journal} {Phys. Rev. Lett.}\ }\textbf {\bibinfo {volume} {114}},\ \bibinfo {pages} {147201} (\bibinfo {year} {2015})}\BibitemShut {NoStop}%
\bibitem [{\citenamefont {Wang}\ \emph {et~al.}(2017{\natexlab{b}})\citenamefont {Wang}, \citenamefont {Reschke}, \citenamefont {H\"uvonen}, \citenamefont {Do}, \citenamefont {Choi}, \citenamefont {Gensch}, \citenamefont {Nagel}, \citenamefont {R\~o\ om},\ and\ \citenamefont {Loidl}}]{PhysRevLett.119.227202}%
  \BibitemOpen
  \bibfield  {author} {\bibinfo {author} {\bibfnamefont {Z.}~\bibnamefont {Wang}}, \bibinfo {author} {\bibfnamefont {S.}~\bibnamefont {Reschke}}, \bibinfo {author} {\bibfnamefont {D.}~\bibnamefont {H\"uvonen}}, \bibinfo {author} {\bibfnamefont {S.-H.}\ \bibnamefont {Do}}, \bibinfo {author} {\bibfnamefont {K.-Y.}\ \bibnamefont {Choi}}, \bibinfo {author} {\bibfnamefont {M.}~\bibnamefont {Gensch}}, \bibinfo {author} {\bibfnamefont {U.}~\bibnamefont {Nagel}}, \bibinfo {author} {\bibfnamefont {T.}~\bibnamefont {R\~o\ om}},\ and\ \bibinfo {author} {\bibfnamefont {A.}~\bibnamefont {Loidl}},\ }\href {https://doi.org/10.1103/PhysRevLett.119.227202} {\bibfield  {journal} {\bibinfo  {journal} {Phys. Rev. Lett.}\ }\textbf {\bibinfo {volume} {119}},\ \bibinfo {pages} {227202} (\bibinfo {year} {2017}{\natexlab{b}})}\BibitemShut {NoStop}%
\bibitem [{\citenamefont {Sahasrabudhe}\ \emph {et~al.}(2020)\citenamefont {Sahasrabudhe}, \citenamefont {Kaib}, \citenamefont {Reschke}, \citenamefont {German}, \citenamefont {Koethe}, \citenamefont {Buhot}, \citenamefont {Kamenskyi}, \citenamefont {Hickey}, \citenamefont {Becker}, \citenamefont {Tsurkan}, \citenamefont {Loidl}, \citenamefont {Do}, \citenamefont {Choi}, \citenamefont {Gr\"uninger}, \citenamefont {Winter}, \citenamefont {Wang}, \citenamefont {Valent\'{\i}},\ and\ \citenamefont {van Loosdrecht}}]{PhysRevB.101.140410}%
  \BibitemOpen
  \bibfield  {author} {\bibinfo {author} {\bibfnamefont {A.}~\bibnamefont {Sahasrabudhe}}, \bibinfo {author} {\bibfnamefont {D.~A.~S.}\ \bibnamefont {Kaib}}, \bibinfo {author} {\bibfnamefont {S.}~\bibnamefont {Reschke}}, \bibinfo {author} {\bibfnamefont {R.}~\bibnamefont {German}}, \bibinfo {author} {\bibfnamefont {T.~C.}\ \bibnamefont {Koethe}}, \bibinfo {author} {\bibfnamefont {J.}~\bibnamefont {Buhot}}, \bibinfo {author} {\bibfnamefont {D.}~\bibnamefont {Kamenskyi}}, \bibinfo {author} {\bibfnamefont {C.}~\bibnamefont {Hickey}}, \bibinfo {author} {\bibfnamefont {P.}~\bibnamefont {Becker}}, \bibinfo {author} {\bibfnamefont {V.}~\bibnamefont {Tsurkan}}, \bibinfo {author} {\bibfnamefont {A.}~\bibnamefont {Loidl}}, \bibinfo {author} {\bibfnamefont {S.~H.}\ \bibnamefont {Do}}, \bibinfo {author} {\bibfnamefont {K.~Y.}\ \bibnamefont {Choi}}, \bibinfo {author} {\bibfnamefont {M.}~\bibnamefont {Gr\"uninger}}, \bibinfo {author} {\bibfnamefont {S.~M.}\ \bibnamefont {Winter}}, \bibinfo {author} {\bibfnamefont
  {Z.}~\bibnamefont {Wang}}, \bibinfo {author} {\bibfnamefont {R.}~\bibnamefont {Valent\'{\i}}},\ and\ \bibinfo {author} {\bibfnamefont {P.~H.~M.}\ \bibnamefont {van Loosdrecht}},\ }\href {https://doi.org/10.1103/PhysRevB.101.140410} {\bibfield  {journal} {\bibinfo  {journal} {Phys. Rev. B}\ }\textbf {\bibinfo {volume} {101}},\ \bibinfo {pages} {140410} (\bibinfo {year} {2020})}\BibitemShut {NoStop}%
\bibitem [{\citenamefont {Pilch}\ \emph {et~al.}(2023)\citenamefont {Pilch}, \citenamefont {Peedu}, \citenamefont {Bera}, \citenamefont {Yusuf}, \citenamefont {Nagel}, \citenamefont {R\~o\ om},\ and\ \citenamefont {Wang}}]{PhysRevB.108.L140406}%
  \BibitemOpen
  \bibfield  {author} {\bibinfo {author} {\bibfnamefont {P.}~\bibnamefont {Pilch}}, \bibinfo {author} {\bibfnamefont {L.}~\bibnamefont {Peedu}}, \bibinfo {author} {\bibfnamefont {A.~K.}\ \bibnamefont {Bera}}, \bibinfo {author} {\bibfnamefont {S.~M.}\ \bibnamefont {Yusuf}}, \bibinfo {author} {\bibfnamefont {U.}~\bibnamefont {Nagel}}, \bibinfo {author} {\bibfnamefont {T.}~\bibnamefont {R\~o\ om}},\ and\ \bibinfo {author} {\bibfnamefont {Z.}~\bibnamefont {Wang}},\ }\href {https://doi.org/10.1103/PhysRevB.108.L140406} {\bibfield  {journal} {\bibinfo  {journal} {Phys. Rev. B}\ }\textbf {\bibinfo {volume} {108}},\ \bibinfo {pages} {L140406} (\bibinfo {year} {2023})}\BibitemShut {NoStop}%
\bibitem [{\citenamefont {Xiang}\ \emph {et~al.}(2023)\citenamefont {Xiang}, \citenamefont {Dhakal}, \citenamefont {Ozerov}, \citenamefont {Jiang}, \citenamefont {Mou}, \citenamefont {Ozarowski}, \citenamefont {Huang}, \citenamefont {Zhou}, \citenamefont {Fang}, \citenamefont {Winter}, \citenamefont {Jiang},\ and\ \citenamefont {Smirnov}}]{PhysRevLett.131.076701}%
  \BibitemOpen
  \bibfield  {author} {\bibinfo {author} {\bibfnamefont {L.}~\bibnamefont {Xiang}}, \bibinfo {author} {\bibfnamefont {R.}~\bibnamefont {Dhakal}}, \bibinfo {author} {\bibfnamefont {M.}~\bibnamefont {Ozerov}}, \bibinfo {author} {\bibfnamefont {Y.}~\bibnamefont {Jiang}}, \bibinfo {author} {\bibfnamefont {B.~S.}\ \bibnamefont {Mou}}, \bibinfo {author} {\bibfnamefont {A.}~\bibnamefont {Ozarowski}}, \bibinfo {author} {\bibfnamefont {Q.}~\bibnamefont {Huang}}, \bibinfo {author} {\bibfnamefont {H.}~\bibnamefont {Zhou}}, \bibinfo {author} {\bibfnamefont {J.}~\bibnamefont {Fang}}, \bibinfo {author} {\bibfnamefont {S.~M.}\ \bibnamefont {Winter}}, \bibinfo {author} {\bibfnamefont {Z.}~\bibnamefont {Jiang}},\ and\ \bibinfo {author} {\bibfnamefont {D.}~\bibnamefont {Smirnov}},\ }\href {https://doi.org/10.1103/PhysRevLett.131.076701} {\bibfield  {journal} {\bibinfo  {journal} {Phys. Rev. Lett.}\ }\textbf {\bibinfo {volume} {131}},\ \bibinfo {pages} {076701} (\bibinfo {year} {2023})}\BibitemShut {NoStop}%
\bibitem [{\citenamefont {Jiang}\ \emph {et~al.}(2023)\citenamefont {Jiang}, \citenamefont {White},\ and\ \citenamefont {Chernyshev}}]{DMRG_J1_J3}%
  \BibitemOpen
  \bibfield  {author} {\bibinfo {author} {\bibfnamefont {S.}~\bibnamefont {Jiang}}, \bibinfo {author} {\bibfnamefont {S.~R.}\ \bibnamefont {White}},\ and\ \bibinfo {author} {\bibfnamefont {A.~L.}\ \bibnamefont {Chernyshev}},\ }\href {https://doi.org/10.1103/PhysRevB.108.L180406} {\bibfield  {journal} {\bibinfo  {journal} {Phys. Rev. B}\ }\textbf {\bibinfo {volume} {108}},\ \bibinfo {pages} {L180406} (\bibinfo {year} {2023})}\BibitemShut {NoStop}%
\bibitem [{\citenamefont {Miao}\ \emph {et~al.}(2024)\citenamefont {Miao}, \citenamefont {Jin}, \citenamefont {Yao}, \citenamefont {Chen}, \citenamefont {Koda}, \citenamefont {Tan}, \citenamefont {Xie}, \citenamefont {Ji}, \citenamefont {Kamiyama},\ and\ \citenamefont {Li}}]{NCSO_residual_fluctuations}%
  \BibitemOpen
  \bibfield  {author} {\bibinfo {author} {\bibfnamefont {P.}~\bibnamefont {Miao}}, \bibinfo {author} {\bibfnamefont {X.}~\bibnamefont {Jin}}, \bibinfo {author} {\bibfnamefont {W.}~\bibnamefont {Yao}}, \bibinfo {author} {\bibfnamefont {Y.}~\bibnamefont {Chen}}, \bibinfo {author} {\bibfnamefont {A.}~\bibnamefont {Koda}}, \bibinfo {author} {\bibfnamefont {Z.}~\bibnamefont {Tan}}, \bibinfo {author} {\bibfnamefont {W.}~\bibnamefont {Xie}}, \bibinfo {author} {\bibfnamefont {W.}~\bibnamefont {Ji}}, \bibinfo {author} {\bibfnamefont {T.}~\bibnamefont {Kamiyama}},\ and\ \bibinfo {author} {\bibfnamefont {Y.}~\bibnamefont {Li}},\ }\href {https://doi.org/10.1103/PhysRevB.109.134431} {\bibfield  {journal} {\bibinfo  {journal} {Phys. Rev. B}\ }\textbf {\bibinfo {volume} {109}},\ \bibinfo {pages} {134431} (\bibinfo {year} {2024})}\BibitemShut {NoStop}%
\bibitem [{\citenamefont {Yao}\ and\ \citenamefont {Li}(2020)}]{yao2020ferrimagnetism}%
  \BibitemOpen
  \bibfield  {author} {\bibinfo {author} {\bibfnamefont {W.}~\bibnamefont {Yao}}\ and\ \bibinfo {author} {\bibfnamefont {Y.}~\bibnamefont {Li}},\ }\href {https://doi.org/10.1103/PhysRevB.101.085120} {\bibfield  {journal} {\bibinfo  {journal} {Phys. Rev. B}\ }\textbf {\bibinfo {volume} {101}},\ \bibinfo {pages} {085120} (\bibinfo {year} {2020})}\BibitemShut {NoStop}%
\bibitem [{\citenamefont {Lampen-Kelley}\ \emph {et~al.}(2018)\citenamefont {Lampen-Kelley}, \citenamefont {Rachel}, \citenamefont {Reuther}, \citenamefont {Yan}, \citenamefont {Banerjee}, \citenamefont {Bridges}, \citenamefont {Cao}, \citenamefont {Nagler},\ and\ \citenamefont {Mandrus}}]{Mandrus_PRB_2018}%
  \BibitemOpen
  \bibfield  {author} {\bibinfo {author} {\bibfnamefont {P.}~\bibnamefont {Lampen-Kelley}}, \bibinfo {author} {\bibfnamefont {S.}~\bibnamefont {Rachel}}, \bibinfo {author} {\bibfnamefont {J.}~\bibnamefont {Reuther}}, \bibinfo {author} {\bibfnamefont {J.-Q.}\ \bibnamefont {Yan}}, \bibinfo {author} {\bibfnamefont {A.}~\bibnamefont {Banerjee}}, \bibinfo {author} {\bibfnamefont {C.~A.}\ \bibnamefont {Bridges}}, \bibinfo {author} {\bibfnamefont {H.~B.}\ \bibnamefont {Cao}}, \bibinfo {author} {\bibfnamefont {S.~E.}\ \bibnamefont {Nagler}},\ and\ \bibinfo {author} {\bibfnamefont {D.}~\bibnamefont {Mandrus}},\ }\href {https://doi.org/10.1103/PhysRevB.98.100403} {\bibfield  {journal} {\bibinfo  {journal} {Phys. Rev. B}\ }\textbf {\bibinfo {volume} {98}},\ \bibinfo {pages} {100403} (\bibinfo {year} {2018})}\BibitemShut {NoStop}%
\bibitem [{\citenamefont {Rau}\ \emph {et~al.}(2016)\citenamefont {Rau}, \citenamefont {Lee},\ and\ \citenamefont {Kee}}]{HYKee_ARCMP_2016}%
  \BibitemOpen
  \bibfield  {author} {\bibinfo {author} {\bibfnamefont {J.~G.}\ \bibnamefont {Rau}}, \bibinfo {author} {\bibfnamefont {E.~K.-H.}\ \bibnamefont {Lee}},\ and\ \bibinfo {author} {\bibfnamefont {H.-Y.}\ \bibnamefont {Kee}},\ }\href {https://doi.org/https://doi.org/10.1146/annurev-conmatphys-031115-011319} {\bibfield  {journal} {\bibinfo  {journal} {Annual Review of Condensed Matter Physics}\ }\textbf {\bibinfo {volume} {7}},\ \bibinfo {pages} {195} (\bibinfo {year} {2016})}\BibitemShut {NoStop}%
\bibitem [{\citenamefont {Maksimov}\ and\ \citenamefont {Chernyshev}(2022)}]{Chernyshev_PRB_2022}%
  \BibitemOpen
  \bibfield  {author} {\bibinfo {author} {\bibfnamefont {P.~A.}\ \bibnamefont {Maksimov}}\ and\ \bibinfo {author} {\bibfnamefont {A.~L.}\ \bibnamefont {Chernyshev}},\ }\href {https://doi.org/10.1103/PhysRevB.106.214411} {\bibfield  {journal} {\bibinfo  {journal} {Phys. Rev. B}\ }\textbf {\bibinfo {volume} {106}},\ \bibinfo {pages} {214411} (\bibinfo {year} {2022})}\BibitemShut {NoStop}%
\end{thebibliography}%

\pagebreak
\pagebreak
\pagebreak

\widetext

\begin{center}
\textbf{Supplementary Information for "Field-induced spin continuum in twin-free Na$_3$Co$_2$SbO$_6$ revealed by magneto-THz spectroscopy"}
\end{center}

\renewcommand{\theequation}{S\arabic{equation}}
\renewcommand{\thetable}{S\arabic{table}}
\renewcommand{\thefigure}{S\arabic{figure}}

\setcounter{figure}{0}
\section{Single crystal growth}

Single crystals of NCSO were synthesized using a flux method similar to that employed for growing its sister compound Na$_2$Co$_2$TeO$_6$ \cite{li2022giant,yao2020ferrimagnetism}. The high-quality, twin-free single crystals, identified through Raman spectroscopy, exhibit a flaky hexagonal shape and a dark purple color \cite{li2022giant}. The twin-free nature of the single crystals was also verified through single-crystal X-ray diffraction, neutron diffraction, and magnetization experiments \cite{li2022giant, Gu_NCSO_double_Q}. The size of the obtained single crystals was approximately 2 mm $\times$ 2 mm $\times$ 0.2 mm.

\section{Magnetization measurements}
\begin{figure*}[htbp]
\includegraphics[clip, width=0.9\textwidth]{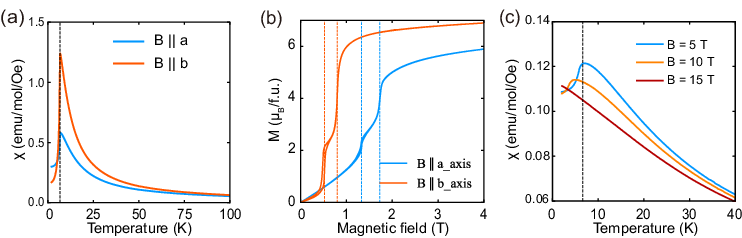}\\[1pt] 
\caption{(a) The magnetic susceptibility of NCSO was measured in zero-field-cooling mode at $B$ = 0.1 T for magnetic fields along the $\bf{a}$ axis and $\bf{a}$ axis. The antiferromagnetic phase transition is observed at $T_\mathrm{N} = 6.6$ K. (b) Magnetization versus field data features two transitions with external fields along the $\bf{a}$ and $\bf{b}$ axis. The critical fields are determined as $B_{c1}$ = 1.3 T, $B_{c2}$ = 1.7 T along $\bf{a}$ axis; $B_{c1}$ = 0.5 T, $B_{c2}$ = 0.8 T along $\bf{b}$ axis. (c) The temperature dependence of the magnetic susceptibility of NCSO at selected magnetic fields with external fields along the $\bf{c}$ axis.}

\label{fig:S1}
\end{figure*}

Magnetic measurements (DC susceptibility) were carried out using a Quantum Design Physical Property Measurement System (PPMS). In a typical temperature-sweep experiment, the sample was cooled to 2 K under zero-field-cooled (ZFC) conditions, and data were collected upon warming from 2 to 100 K in an applied in-plane magnetic field of 1000 Oe. Below the Néel temperature ($T_\mathrm{N} = 6.6$ K), the compound adopts an in-plane antiferromagnetic spin structure, as clearly evidenced by the magnetic susceptibility data shown in Fig.~\ref{fig:S1}(a). The susceptibility exhibits pronounced anisotropy between the $\bf{a}$ and $\bf{b}$ axes, consistent with the findings reported in Ref.~\cite{li2022giant}. Upon increasing the magnetic field, the magnetization along both the $\bf{a}$ and $\bf{b}$ axes displays two successive transitions. The corresponding critical fields, $B_{c1}$ and $B_{c2}$, were determined from the derivatives of the magnetization curves to be (1.3 T, 1.7 T) for the $\bf{a}$ axis and (0.5 T, 0.8 T) for the $\bf{b}$ axis, as shown in Fig.~\ref{fig:S1}(b). In contrast, when the magnetic field was applied out of the plane, no phase transitions were observed up to 15 T [Fig.~\ref{fig:S1}(c)].

\section{Time-domain terahertz spectroscopy}

Time-domain THz transmission spectra were measured using a home-built spectroscopy system equipped with an Oxford Spectromag magnet \cite{shi2018field}. THz pulses, with a bandwidth ranging from 0.2 THz to 2 THz, were generated by a photoconductive antenna upon illumination with an 800 nm laser. The time-domain signals of the sample and the reference (an empty aperture of the same size) were detected via free-space electro-optic sampling in ZnTe.

The complex transmission of the sample is given by the relation:

\begin{equation}
\widetilde{T}=\frac{4\widetilde{n}}{(1+\widetilde{n})^2}exp(\frac{i\omega d}{c}(\widetilde{n}-1))
\label{eqS1}
\end{equation}

where $\widetilde{n}$ is the complex refractive index of the sample, $\omega$ is the angular frequency, $d$ is the sample thickness, and $c$ is the speed of light. The complex equation is solved using a numerical iterative method to obtain the frequency-dependent real and imaginary parts of the complex refractive index $\widetilde{n}$. In this insulating compound, we assume that the dielectric contribution is small compared to the magnetic susceptibility $\widetilde{\chi}$, allowing the approximation $\widetilde{n} \approx n(1 + \widetilde{\chi}/2)$. The imaginary part of the magnetic susceptibility is approximately proportional to the imaginary of the refractive index, $\chi_2 (\omega)= 2 \kappa /n$ \cite{shi2018field}.

\begin{figure*}[htbp]
\includegraphics[clip, width=0.9\textwidth]{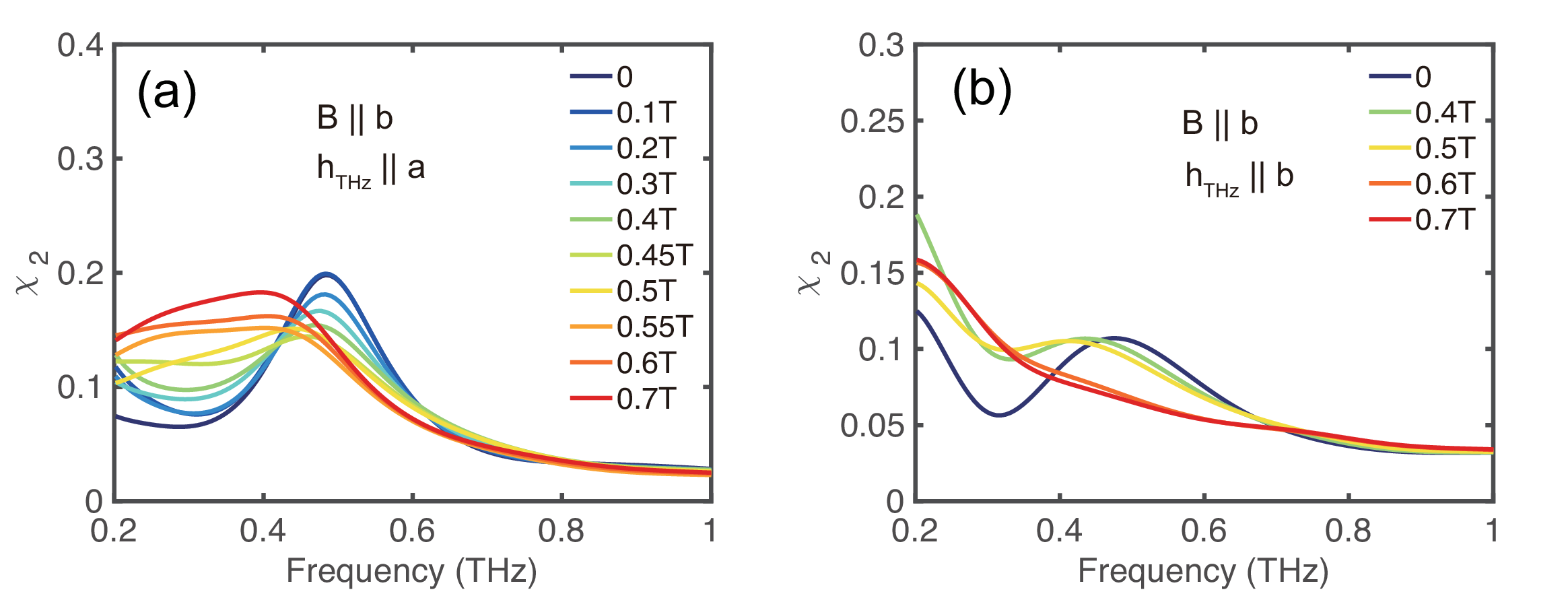}\\[1pt] 
\caption{$\chi_2$($\omega$) at 2 K at different magnetic fields with the magnetic field along the $\bf{b}$ axis. (a)  $\bf{h} _\mathbf{THz} \perp \bf{b}$. (b) $\bf{h} _\mathbf{THz} \perp \bf{b}$. 
\label{fig:S3}}
\end{figure*}

\section{Magnetic susceptibility with $\mathbf{B} \parallel \mathbf{b}$}

The magnetic susceptibility $\chi_2(\omega)$ measured at 2 K with the external magnetic field applied along the $b$-axis is shown in Fig.~\ref{fig:S3}. Although strong magnetic anisotropy is present, the observed behavior is similar to that for $\mathbf{B} \parallel \mathbf{a}$. As the magnetic field increases, the magnon mode is suppressed around 0.5 T, and a low-energy continuum dominates the spectrum above this field. The continuum persists until the spins are fully polarized by the external field. The critical fields for $\mathbf{B} \parallel \mathbf{b}$ are 0.5 T and 0.8 T, which are lower than those for $\mathbf{B} \parallel \mathbf{a}$.

\begin{figure*}[htbp]
\centering
\includegraphics[clip, width=0.9\textwidth]{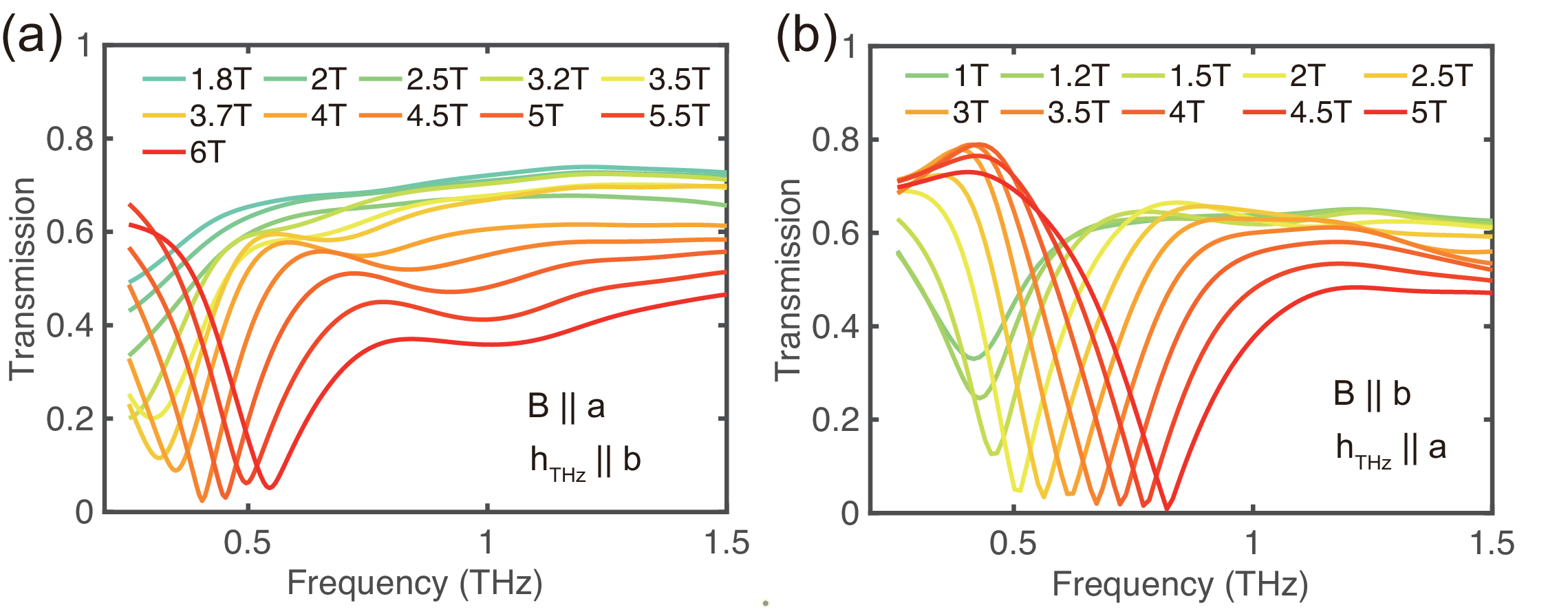}

\caption{Transmission spectrum in the field-polarized state with $\mathbf{h}_{\mathrm{THz}} \perp \mathbf{B}$. The spectrum is dominated by the sharp single-magnon mode. (a) $\mathbf{B} \parallel \mathbf{a}$. (b) $\mathbf{B} \parallel \mathbf{b}$.}
\label{fig:S4}
\end{figure*}

\section{Terahertz transmission in the field-polarized state}\label{ap1}

\begin{figure*}[htbp]
\centering
\includegraphics[clip, width=0.9\textwidth]{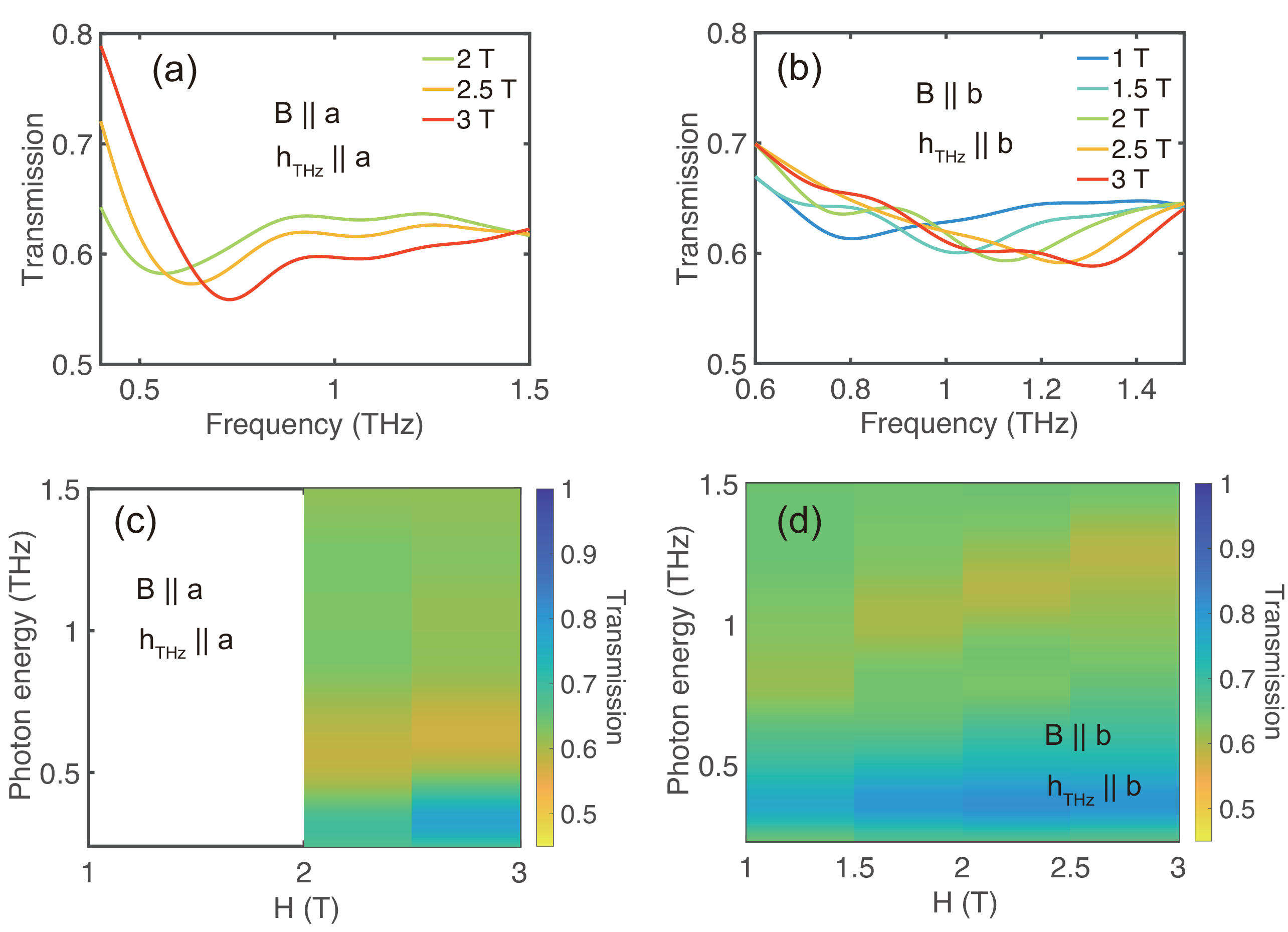}

\caption{Transmission spectrum in longitudinal polarization ($\mathbf{h}_{\mathrm{THz}} \parallel \mathbf{B}$).  (a)(b) Transmission with $\mathbf{h}_{\mathrm{THz}} \parallel \mathbf{B} \parallel \mathbf{a}$ and $\mathbf{h}_{\mathrm{THz}} \parallel \mathbf{B} \parallel \mathbf{b}$, respectively. The spectrum is dominated by the broad two-magnon feature. (c)(d) Color plot of the transmission as a function of frequency and magnetic field.}
\label{fig:S5}
\end{figure*}

In the field-polarized state, the excitation spectrum is dominated by single-magnon and two-magnon processes. Single magnons are only detected in transverse polarization ($\bf{h}_{\mathbf{THz}} \perp \mathbf{B}$) due to Zeeman selection rules. The transmission spectrum in this configuration is shown in Fig.~\ref{fig:S4}. In longitudinal polarization ($\bf{h}_{\mathbf{THz}} \parallel \mathbf{B}$), the spectrum is dominated by the two-magnon continuum in the polarized state. The transmission spectrum with $\bf{h}_{\mathbf{THz}} \parallel \mathbf{B}$ in the polarized state is shown in Fig.~\ref{fig:S5}. The two-magnon excitation appears as a broad feature starting from approximately twice the frequency of the single magnons. Similar features are also observed in the transverse configuration, suggesting the spins have not been fully polarized at these fields.

\section{Magnon energy at $\Vect{q} = 0$ in fully polarized states for generic bilinear spin models on the honeycomb lattice for Na$_3$Co$_2$SbO$_6$}

We assume that Na$_3$Co$_2$SbO$_6$ is quasi-two-dimensional and consider a single layer of the Co honeycomb lattice. It is nearly a perfect honeycomb lattice, but the entire crystal and therefore the spin Hamiltonian does not possess exact $D_{6h}$ point group symmetry. Note that spatial point group symmetry operations should be accompanied by the corresponding transformations in spin space due to spin-orbit coupling effects.

We use the orthogonal crystallographic axes along $\Vect{a}$, $\Vect{b}$, and $\Vect{c}^* \propto \Vect{a} \times \Vect{b}$, and denote vector components along these directions by the subscripts $a$, $b$, and $c^*$, respectively. Denote the two sublattices of the honeycomb lattice by $A$ and $B$. The $A$-to-$B$ displacements of the $X$, $Y$, and $Z$ types of nearest-neighbor bonds are given by $-\frac{1}{2}\Vect{a} - \frac{1}{6}\Vect{b}$, $+\frac{1}{2}\Vect{a} - \frac{1}{6}\Vect{b}$, and $\frac{1}{3}\Vect{b}$, respectively.

The reduced point group symmetry for Na$_3$Co$_2$SbO$_6$ is $C_{2h}$ about the $Z$ bond centers, generated by the following two symmetry operations:
\begin{itemize}
\item
Spatial inversion about the $Z$ bond centers, which exchanges the $A$ and $B$ sublattices without acting on the spins. This implies that all bond centers and hexagon centers are inversion centers, so there are no Dzyaloshinskii–Moriya interactions on nearest-neighbor and third-neighbor bonds.

\item
$C_2$ rotation around the $Z$ bonds (i.e., around the $\Vect{b}$ axis), accompanied by the same $C_2$ rotation in spin space ($S_{a} \to -S_{a}$, $S_{b} \to S_{b}$, $S_{c^*} \to -S_{c^*}$). We define the associated rotation matrix as
\[
\mat{R}_{b} =
\begin{pmatrix}
-1 & 0 & 0 \\
0 & 1 & 0 \\
0 & 0 & -1
\end{pmatrix}.
\]
\end{itemize}

The generic bilinear spin model under an external magnetic field $\Vect{\hfield}$ is

\begin{equation}
\ham = \sum_{(ij)} \Vect{S}_i \cdot \mat{J}_{ij} \cdot \Vect{S}_j 
- \mu_B \Vect{\hfield} \cdot \sum_i \mat{G}_i \cdot \Vect{S}_i,
\label{eqS2}
\end{equation}

where the first sum runs over all distinct bonds $(ij)$ with a $3 \times 3$ bilinear coupling matrix $\mat{J}_{ij}$, and the second sum runs over all spin sites $i$ with $g$-factor tensor (referred to as the $g$-tensor hereafter) $\mat{G}_i$.

The space group symmetry requires that all $g$-tensors are identical, i.e., $\mat{G}_i = \mat{G}$, and satisfy $\mat{R}_{b} \cdot \mat{G} \cdot \mat{R}_b^{-1} = \mat{G}$, meaning that $\Vect{b}$ must be a principal axis of the $g$-tensor. To simplify later discussions, we assume that $\mat{G}$ is diagonal in the crystallographic basis, and that the $g$-factors along the $\Vect{a}$, $\Vect{b}$, and $\Vect{c}^*$ axes are $g_{a}$, $g_{b}$, and $g_{c^*}$, respectively. (This assumption differs from that in Ref.~\cite{Mandrus_PRB_2018}.)

In the following, we consider only $\Vect{q} = 0$ classical spin configurations and magnon excitations. Their energies are fully determined by the $\Vect{q} = 0$ spin interaction matrices,
\[
\mat{J}_{\ell \ell'} = \sum_{j \in \ell'} \mat{J}_{ij},
\]
where site $i$ belongs to sublattice $\ell = A, B$, and the sum runs over all sites $j$ in sublattice $\ell' = A, B$.

The condition $\mat{J}_{ji} = \transpose{\mat{J}_{ij}}$ and the $C_{2h}$ point group symmetry impose the following constraints:
\[
\mat{J}_{\ell' \ell} = \transpose{\mat{J}_{\ell \ell'}}, \quad
\mat{J}_{\ell' \ell} = \mat{J}_{\ell \ell'}, \quad
\mat{R}_{b} \cdot \mat{J}_{\ell \ell'} \cdot \mat{R}_{b}^{-1} = \mat{J}_{\ell \ell'}.
\]
Therefore, all $\mat{J}_{\ell \ell'}$ are real symmetric matrices with $\Vect{b}$ as one of their principal axes.

\subsection{Magnon energies for diagonal $\Vect{q}=0$ interactions under crystallographic axes}
\label{subsection:diagonal_case}

First consider the simplest case in which the interaction matrices are diagonal:
\[
\mat{J}_{\ell \ell'} = \begin{pmatrix}
\Jeff_{\ell \ell', a} & 0 & 0 \\
0 & \Jeff_{\ell \ell', b} & 0 \\
0 & 0 & \Jeff_{\ell \ell', c^*}
\end{pmatrix}
\]
in the crystallographic basis. Then, if the external field is applied along a crystallographic axis, the fully polarized moments will align along that axis as well.

For a large magnetic field $\Vect{\hfield} = (\hfield, 0, 0)$ and spins fully polarized along the $\Vect{a}$ axis,  
the linear spin wave Hamiltonian at $\Vect{q} = 0$ takes the form:
\begin{equation}
\hat{H}_{\text{LSW, polarized along }\Vect{a}}(\Vect{q}=0)=\frac{1}{2}
\begin{pmatrix}
\hat{b}^\dagger_{A} & \hat{b}^\dagger_{B} & \hat{b}^\nd_{A} & \hat{b}^\nd_{B}
\end{pmatrix}
\cdot 
\mat{H}_{a}({\Vect{q}=0})
\cdot 
\begin{pmatrix}
\hat{b}^\dagger_{A} \\ \hat{b}^\dagger_{B} \\ \hat{b}^\nd_{A} \\ \hat{b}^\nd_{B}
\end{pmatrix}
\end{equation}

Here $\hat{b}_{\ell}$ is the annihilation operator for $\Vect{q}=0$ Holstein-Primakoff boson mode on sublattice $\ell=A,B$.
The $4\times 4$ real symmetric matrix $\mat{H}_{a}({\Vect{q}=0})$ is
\begin{equation}
\begin{array}{rl}
\mat{H}_{a}({\Vect{q}=0})
=
& 
[\mu_B \hfield g_{a}-S\cdot(\Jeff_{AA,a}+\Jeff_{AA,a+}+\Jeff_{AB,a})]\cdot \idmat_{4\times 4}
\\
&
-S \begin{pmatrix}
0 & -\Jeff_{AB,a+} & \Jeff_{AA,a-} & \Jeff_{AB,a-}\\
-\Jeff_{AB,a+} & 0 & \Jeff_{AB,a-} & \Jeff_{AA,a-}\\
\Jeff_{AA,a-} & \Jeff_{AB,a-} & 0 & -\Jeff_{AB,a+}\\
\Jeff_{AB,a-} & \Jeff_{AA,a-} & -\Jeff_{AB,a+} & 0\\
\end{pmatrix}
\end{array}
\end{equation}

Here $\idmat_{4\times 4}$ is the $4\times 4$ identity matrix,
$\Jeff_{\ell\ell',a\pm}\equiv \frac{1}{2}(\Jeff_{\ell\ell',b}\pm \Jeff_{\ell\ell',c^*})$. 
Diagonalize $\mat{H}_{a}(\Vect{q}=0)$ by $SU(2,2)$ Bogoliubov transformation
produces the two $\Vect{q}=0$ magnon energies,
\begin{equation}
\epsilon_{a,\pm}(\Vect{q}=0)
=
\sqrt{
[
\mu_B \hfield g_{a}-S\cdot(\Jeff_{AA,a}+\Jeff_{AB,a}-\Jeff_{AA,a+}\mp \Jeff_{AB,a+})
]^2
-S^2\cdot (\Jeff_{AA,a-}\pm \Jeff_{AB,a-})^2
}
\end{equation}

The top and bottom signs correspond to the symmetric ($\hat{b}_{A} + \hat{b}_{B}$) and antisymmetric ($\hat{b}_{A} - \hat{b}_{B}$) magnon modes, respectively. Our THz experiments exclusively probe the symmetric modes, with energy given by:

\begin{equation}
\epsilon_{a,+}(\Vect{q}=0)
=
\sqrt{
[
\mu_B \hfield g_{a}-S\cdot(\Jeff_{+,a}-\frac{\Jeff_{+,b}+\Jeff_{+,c^*}}{2})
]^2
-S^2\cdot (\frac{\Jeff_{+,b}- \Jeff_{+,c^*}}{2})^2
}
\end{equation}

and the parameters $\Jeff_{+,d}\equiv \Jeff_{AA,d}+\Jeff_{AB,d}$ for $d=a,b,c^*$ 
are related to the Curie-Weiss temperatures of high temperature magnetic susceptibility 
$\chi_{d}(T)\sim \frac{\text{constant}}{T-\CWtemp_{d}}$ 
along axis $d$ by 
$k_B\CWtemp_{d}=-\frac{S(S+1)}{3}\Jeff_{+,d}=-\frac{1}{4}\Jeff_{+,d}$ for $S=\frac{1}{2}$ systems\cite{Mandrus_PRB_2018}
.

Finally we have the magnon energies probed by THz experiments in terms of Curie-Weiss temperatures:
\begin{equation}
\label{equ:epsilon_polarized}
\epsilon_{a,+}(\Vect{q}=0)
=
\sqrt{
(\mu_B \hfield g_{a}+2k_B\CWtemp_{a}-2k_B\CWtemp_{b})
(\mu_B \hfield g_{a}+2k_B\CWtemp_{a}-2k_B\CWtemp_{c^*})
}
\end{equation}
For very large $\hfield$, 
\begin{equation}
\epsilon_{a,+}(\Vect{q}=0)
\approx
\mu_B \hfield g_{a}+k_B(2\CWtemp_{a}-\CWtemp_{b}-\CWtemp_{c^*})
-\frac{k_B^2(\CWtemp_{b}-\CWtemp_{c^*})^2}{2\mu_B \hfield g_a}+O(\hfield^{-2})
\end{equation}

The classical saturation field $\hfield_{\text{sat},\Vect{a}}$ must ensure that 
both $\epsilon_{a,\pm}(\Vect{q}=0)$ are real positive, 
namely $\mu_B\hfield_{\text{sat},\Vect{a}}g_{a}
\geq -2k_B\CWtemp_{a}+2\max(k_B|\CWtemp_{b}|, k_B|\CWtemp_{c^*}|)$. 
Note that this condition is necessary but not sufficient,
because the moments may have $\Vect{q}\neq 0$ components perpendicular to $\Vect{a}$.

For field and spins polarized along other crystallographic axes, we just need to cyclically permute the $a,b,c^*$ subscripts in the above formulas.

\subsection{The anisotropic nearest-neighbor Kitaev-Heisenberg-$\Gamma$ model}
The crystallographic axes $\Vect{a}$ and $\Vect{b}$ are not along the $x,y,z$ axes of the Kitaev interactions. 
For honeycomb lattice formed by ideal edge-sharing oxygen octahedra, 
$\Vect{a}$, $\Vect{b}$ and $\Vect{c}^*\propto \Vect{a}\times\Vect{b}$ would be along $(1,1,-2)$, $(-1,1,0)$ and $(1,1,1)$ directions of the Kitaev spin axes, respectively.

For the anisotropic nearest-neighbor Kitaev-Heisenberg-$\Gamma$ model
\cite{HYKee_ARCMP_2016}, 
the nearest-neighbor coupling matrices \emph{under the Kitaev spin axes} are
\begin{equation}
\mat{J}^{\KitaevAxis}_X=\begin{pmatrix}
J_X+K_X & \Gamma''_X & \Gamma'_X\\
\Gamma''_X & J_X & \Gamma_X\\
\Gamma'_X & \Gamma_X & J_X
\end{pmatrix},\ 
\mat{J}^{\KitaevAxis}_Y=\begin{pmatrix}
J_Y & \Gamma''_Y & \Gamma_Y\\
\Gamma''_Y & J_Y+K_Y & \Gamma'_Y\\
\Gamma_Y & \Gamma'_Y & J_Y
\end{pmatrix},\ 
\mat{J}^{\KitaevAxis}_Z=\begin{pmatrix}
J_Z & \Gamma_Z & \Gamma''_Z\\
\Gamma_Z & J_Z & \Gamma'_Z\\
\Gamma''_Z & \Gamma'_Z & J_Z+K_Z
\end{pmatrix},
\end{equation}
for the $X$,$Y$,$Z$ types of bonds, respectively.
Here superscript $\KitaevAxis$ denotes quantities under the Kitaev spin axes.
$J_\alpha$ are Heisenberg couplings, 
$K_\alpha$ are Kitaev interactions,
$\Gamma_\alpha$ and $\Gamma'_\alpha$ and $\Gamma''_\alpha$ are symmetric off-diagonal interactions,
for $\alpha=X,Y,Z$.

Note that the $C_2$ symmetry along $\Vect{b}$ axis is accompanied by 
$C_2$ rotation in spin space ($S_x\to -S_y$, $S_y\to -S_x$, $S_z\to -S_z$),
and implies that $J_Y=J_X$, $K_Y=K_X$, $\Gamma_Y=\Gamma_X$, $\Gamma'_Y=\Gamma'_X$, $\Gamma''_Y=\Gamma''_X$, 
and $\Gamma''_Z=\Gamma'_Z$.

Rotate the spin space axes to the crystallographic axes. 
The results for isotropic nearest-neighbor $K-J-\Gamma$ model under crystallographic axes have been extensively discussed before
\cite{Chernyshev_PRB_2022}
and can be used to check the following results.

Define the orthogonal matrix for the axes change
$\mat{C}=\begin{pmatrix}
\frac{1}{\sqrt{6}} & \frac{1}{\sqrt{6}} & -\frac{2}{\sqrt{6}}\\
-\frac{1}{\sqrt{2}} & \frac{1}{\sqrt{2}} & 0\\
\frac{1}{\sqrt{3}} & \frac{1}{\sqrt{3}} & \frac{1}{\sqrt{3}}
\end{pmatrix}$.
Its rows from top to bottom are the unit vectors along $\Vect{a}$,$\Vect{b}$,$\Vect{c}^*$ directions, respectively.
The nearest-neighbor coupling matrices under crystallographic axes 
$\mat{J}_{\alpha}=\mat{C}\cdot \mat{J}^{\KitaevAxis}_{\alpha}\cdot \mat{C}^{-1}$
are
\begin{subequations}
\begin{align}
\mat{J}_{X} & = 
(J_X-\Gamma_X)\idmat_{3\times 3}+3\Gamma_X\cdot \mat{M}_{ 3,3}
+K_X \cdot \mat{M}_{\subrot,1,1}
+\deltagamma'_X\cdot \mat{M}_{\subrot,1,3}
+\deltagamma''_X\cdot \mat{M}_{\subrot,1,2}
\\
\mat{J}_{Y} & = 
(J_Y-\Gamma_Y)\idmat_{3\times 3}+3\Gamma_Y\cdot \mat{M}_{ 3,3}
+K_Y\cdot \mat{M}_{\subrot,2,2}
+\deltagamma'_Y\cdot \mat{M}_{\subrot,2,3}
+\deltagamma''_Y\cdot \mat{M}_{\subrot,1,2}
\\
\mat{J}_{Z} & = 
(J_Z-\Gamma_Z)\idmat_{3\times 3}+3\Gamma_Z\cdot \mat{M}_{ 3,3}
+K_Z \cdot \mat{M}_{\subrot,3,3}
+\deltagamma'_Z\cdot \mat{M}_{\subrot,2,3}
+\deltagamma''_Z\cdot \mat{M}_{\subrot,1,3}
\end{align}
\end{subequations}

Here $\idmat_{3\times 3}$ is the $3\times 3$ identity matrix, 
$\deltagamma'_\alpha\equiv\Gamma'_\alpha-\Gamma_\alpha$
and $\deltagamma''_\alpha\equiv\Gamma''_\alpha-\Gamma_\alpha$ 
for $\alpha=X,Y,Z$.
Note that if Kitaev interaction $K_\alpha=0$ and $\deltagamma'_\alpha=\deltagamma''_\alpha=0$,
the interactions on bond $\alpha$ are actually XXZ spin interactions under crystallographic axes.
The real symmetric matrices $\mat{M}_{i,j}$ have entries ``$1$'' on the $(i,j)$ and $(j,i)$ elements 
and entries ``$0$'' on other elements. Matrices $\mat{M}_{\subrot,i,j}\equiv \mat{C}\cdot \mat{M}_{i,j}\cdot \mat{C}^{-1}$ are 
\begin{equation*}
\begin{array}{l}
\mat{M}_{\subrot,1,1} 
=\begin{pmatrix}
\frac{1}{6} & -\frac{1}{2\sqrt{3}} & \frac{1}{3\sqrt{2}} \\
-\frac{1}{2\sqrt{3}} & \frac{1}{2} & -\frac{1}{\sqrt{6}} \\
\frac{1}{3\sqrt{2}} & -\frac{1}{\sqrt{6}} & \frac{1}{3}
\end{pmatrix},
\ 
\mat{M}_{\subrot,2,2} 
=\begin{pmatrix}
\frac{1}{6} & \frac{1}{2\sqrt{3}} & \frac{1}{3\sqrt{2}} \\
\frac{1}{2\sqrt{3}} & \frac{1}{2} & \frac{1}{\sqrt{6}} \\
\frac{1}{3\sqrt{2}} & \frac{1}{\sqrt{6}} & \frac{1}{3}
\end{pmatrix},
\ 
\mat{M}_{\subrot,3,3} 
=\begin{pmatrix}
\frac{2}{3} & 0 & -\frac{\sqrt{2}}{3} \\
0 & 0 & 0\\
-\frac{\sqrt{2}}{3} & 0 & \frac{1}{3}
\end{pmatrix},\\
\mat{M}_{\subrot,2,3} 
=\begin{pmatrix}
-\frac{2}{3} & -\frac{1}{\sqrt{3}} & -\frac{1}{3\sqrt{2}}\\
-\frac{1}{\sqrt{3}} & 0 & \frac{1}{\sqrt{6}}\\
-\frac{1}{3\sqrt{2}} & \frac{1}{\sqrt{6}} & \frac{2}{3}
\end{pmatrix},
\ 
\mat{M}_{\subrot,1,3} 
=\begin{pmatrix}
-\frac{2}{3} & \frac{1}{\sqrt{3}} & -\frac{1}{3\sqrt{2}}\\
\frac{1}{\sqrt{3}} & 0 & -\frac{1}{\sqrt{6}}\\
-\frac{1}{3\sqrt{2}} & -\frac{1}{\sqrt{6}} & \frac{2}{3}
\end{pmatrix},
\ 
\mat{M}_{\subrot,1,2}
=\begin{pmatrix}
\frac{1}{3} & 0 & \frac{\sqrt{2}}{3}\\
0 & -1 & 0\\
\frac{\sqrt{2}}{3} & 0 & \frac{2}{3}
\end{pmatrix}
\end{array}
\end{equation*}

The $\Vect{q}=0$ interactions are (see Subsection~\ref{subsection:diagonal_case})
$\mat{J}_{AA}=0$ and 
\begin{equation}
\begin{array}{rl}
\mat{J}_{AB}
\equiv
&
\mat{J}_{X}+\mat{J}_{Y}+\mat{J}_{Z} \\
=
&
(2J_X+J_Z-2\Gamma_X-\Gamma_Z)\idmat_{3\times 3}
+(6\Gamma_X+3\Gamma_Z)\cdot \mat{M}_{3,3}
\\
& 
+K_X\cdot (\mat{M}_{\subrot,1,1}+\mat{M}_{\subrot,2,2})
+K_Z\cdot \mat{M}_{\subrot,3,3}\\
& 
+(\deltagamma'_X+\deltagamma'_Z)\cdot (\mat{M}_{\subrot,1,3}+\mat{M}_{\subrot,2,3})
+\deltagamma''_X\cdot 2\mat{M}_{\subrot,1,2}
\\
=
&
(2J_X+J_Z-2\Gamma_X-\Gamma_Z+K_Z)\idmat_{3\times 3}
\\
&
+
\begin{pmatrix}
-\deltagamma'_X-\deltagamma'_Z+\frac{1}{3}\Delta & 0 & \frac{\sqrt{2}}{3}\Delta\\
0 & 2K_X-2K_Z-\deltagamma'_X-\deltagamma'_Z-\Delta & 0 \\
\frac{\sqrt{2}}{3}\Delta & 0 & 6\Gamma_X+3\Gamma_Z+2\deltagamma^+_X+2\deltagamma'_Z+\frac{2}{3}\Delta
\end{pmatrix}
\end{array}
\end{equation}
Here $\Delta=K_X-K_Z+2\deltagamma''_X-\deltagamma'_X-\deltagamma'_Z$, 
and the $C_{2h}$ symmetry has been exploited so that the $Y$ bond parameters are the same as those on $X$ bonds
and $\deltagamma''_Z=\deltagamma'_Z$.

Then under the assumption that the $g$-tensor $\mat{G}$'s principal axes are crystallographic axes, 
the theoretical Curie-Weiss temperatures are\cite{Mandrus_PRB_2018}.
\begin{subequations}
\begin{align}
\CWtemp_{a}&=-\frac{1}{4k_B}\left [2J_X+J_Z+\frac{2K_Z+K_X}{3}-\frac{4(\Gamma_X+\Gamma'_X+\Gamma'_Z)}{3}+\frac{2\Gamma''_X+\Gamma_Z}{3}\right ]\\
\CWtemp_{b}&=-\frac{1}{4k_B}\left [ 2J_X+J_Z+K_X-(2\Gamma''_X+\Gamma_Z)\right ]\\
\CWtemp_{c^*}&=-\frac{1}{4k_B}\left [2J_X+J_Z+\frac{2K_X+K_Z}{3}+\frac{4(\Gamma_X+\Gamma'_X+\Gamma'_Z)}{3}+\frac{2(2\Gamma''_X+\Gamma_Z)}{3}\right ]
\end{align}
\end{subequations}

To simplify the discussion, we assume that $\Delta=(K_X-K_Z)-(\Gamma_X+\Gamma'_X+\Gamma'_Z)+(2\Gamma''_X+\Gamma_Z)=0$.
Then $\mat{J}_{AB}$ becomes a diagonal matrix (XYZ interaction). 
This would imply that the principal axes of the high-temperature uniform susceptibility tensor
are along the crystallographic axes, 
if the $g$-tensor $\mat{G}$'s principal axes are also along these axes\cite{Mandrus_PRB_2018}.

\subsection{Fitting parameters for the measured magnon energies}
The THz spectroscopy measurements quantitatively determined the field dependence of spin excitations, recording multiple data points for both:
\begin{itemize}
    \item $\epsilon_{a,\text{exp}}(B_{a})$ at various magnetic fields $B_{a} \parallel \mathbf{a}$
    \item $\epsilon_{b,\text{exp}}(B_{b})$ at various magnetic fields $B_{b} \parallel \mathbf{b}$
\end{itemize}
where $B_{a}$ and $B_{b}$ represent the applied magnetic field magnitudes along the respective crystallographic axes.

We employ linear spin wave theory  for fully polarized states (Eq.~\ref{equ:epsilon_polarized}), treating the following as independent parameters:
\begin{itemize}
    \item $g$-factors: $g_{a}$, $g_{b}$
    \item Curie-Weiss temperature differences: 
    $\Theta_{b-a} \equiv \Theta_{b} - \Theta_{a}$,
    $\Theta_{b-c^*} \equiv \Theta_{b} - \Theta_{c^*}$
\end{itemize}

The theoretical magnon energies are then given by:
\begin{subequations}
\begin{align}
\epsilon_{a,+}(B_a) &= \sqrt{(\mu_B B_a g_a - 2k_B\Theta_{b-a})(\mu_B B_a g_a - 2k_B\Theta_{b-a} + 2k_B\Theta_{b-c^*})} \\
\epsilon_{b,+}(B_b) &= \sqrt{(\mu_B B_b g_b + 2k_B\Theta_{b-a})(\mu_B B_b g_b + 2k_B\Theta_{b-c^*})}
\end{align}
\end{subequations}

We minimize 
$\chi\equiv \sum_{B_a} [\epsilon^2_{a,\text{exp}}(\hfield_{a})-\epsilon^2_{a,+}(B_a)]^2
+\sum_{B_b} [\epsilon^2_{b,\text{exp}}(\hfield_{b})-\epsilon^2_{b,+}(B_b)]^2
$
and obtain the following fitting parameters,
\begin{equation}
g_{a}\approx 6.31,\ 
g_{b}\approx 7.14,\ 
\CWtemp_{b}-\CWtemp_{a}\approx 3.66~\mathrm{K},\ 
\CWtemp_{b}-\CWtemp_{c^*}\approx 9.78~\mathrm{K}.
\end{equation}
The comparison between theoretical and experimental results is shown in Fig.~3 (c) in the main text.

With these fitting parameters, 
$\epsilon_{a,+}(B)$ vanishes at $B\approx 1.72~\mathrm{T}$ which is close to the measured polarizing field $1.7~\mathrm{T}$.
And we have the following constraints on model parameters of the anisotropic $K-J-\Gamma$ model, 
\begin{subequations}
\begin{align}
k_B\CWtemp_{b-a}
& = 
\frac{1}{6}\left [(K_Z-K_X)-2(\Gamma_X+\Gamma'_X+\Gamma'_Z)+2(2\Gamma''_X+\Gamma_Z)\right ]
\approx 0.315~\mathrm{meV}
\\
k_B\CWtemp_{b-c^*}
& =
\frac{1}{12}\left [(K_Z-K_X)+4(\Gamma_X+\Gamma'_X+\Gamma'_Z)+5(2\Gamma''_X+\Gamma_Z)\right ]
\approx 0.843~\mathrm{meV}
\end{align}
\end{subequations}

Ref.~\cite{li2022giant} measured some of the Curie-Weiss temperatures and $g$-factors 
by fitting susceptibilities, and reported the following values, 
$\CWtemp_{a}\approx 1.0~\mathrm{K}$, $\CWtemp_{b}\approx 6.8~\mathrm{K}$, 
$g_{a}\approx 6.8$, and $g_{b}\approx 7.3$,
which are roughly consistent with our fitting results.

\end{document}